\documentclass[12pt]{article}
\usepackage{array}
\usepackage{tabularx}
\usepackage{amsmath,amssymb,amsfonts,epsfig,cite,setspace,bigstrut,framed,eufrak}
\usepackage[all]{xy}
\usepackage{color}
\usepackage{pifont}
\usepackage{xcolor}
\usepackage{tikz}
\usepackage[colorlinks=true,urlcolor=blue,anchorcolor=blue,citecolor=blue,filecolor=blue,linkcolor=blue,menucolor=blue,pagecolor=blue,linktocpage=true,pdfproducer=medialab,pdfa=true]{hyperref}	 
\usetikzlibrary{positioning}

\usepackage{mathtools}

\newcommand{\e}{\epsilon}
\newcommand{\dsl}{\pa \kern-0.5em /}

\newcommand{\pa}{\partial}

\newcommand{\bit}{\begin{itemize}}
\newcommand{\eit}{\end{itemize}}

\def\bZ{\mathbb Z}

\newcommand{\bea}{\begin{eqnarray}}
\newcommand{\eea}{\end{eqnarray}}
\newcommand{\be}{\begin{equation}}
\newcommand{\ee}{\end{equation}}

\newcommand{\ba}{\begin{array}}
\newcommand{\ea}{\end{array}}
\def\bZ{\mathbb Z}
\makeatletter \@addtoreset{equation}{section} \makeatother
\renewcommand{\theequation}{\thesection.\arabic{equation}}
\begin{document}

\vskip 0.25in

\newcommand{\todo}[1]{{\bf\color{blue} !! #1 !!}\marginpar{\color{blue}$\Longleftarrow$}}
\newcommand{\comment}[1]{}
\newcommand\T{\rule{0pt}{2.6ex}}
\newcommand\B{\rule[-1.2ex]{0pt}{0pt}}

\newcommand{\CO}{{\cal O}}
\newcommand{\cI}{{\cal I}}
\newcommand{\cM}{{\cal M}}
\newcommand{\cW}{{\cal W}}
\newcommand{\cN}{{\cal N}}
\newcommand{\cR}{{\cal R}}
\newcommand{\cH}{{\cal H}}
\newcommand{\cK}{{\cal K}}
\newcommand{\cT}{{\cal T}}
\newcommand{\cZ}{{\cal Z}}
\newcommand{\cO}{{\cal O}}
\newcommand{\cQ}{{\cal Q}}
\newcommand{\cB}{{\cal B}}
\newcommand{\cC}{{\cal C}}
\newcommand{\cD}{{\cal D}}
\newcommand{\cE}{{\cal E}}
\newcommand{\cF}{{\cal F}}
\newcommand{\cA}{{\cal A}}
\newcommand{\cX}{{\cal X}}
\newcommand{\IA}{\mathbb{A}}
\newcommand{\IP}{\mathbb{P}}
\newcommand{\IQ}{\mathbb{Q}}
\newcommand{\IH}{\mathbb{H}}
\newcommand{\IR}{\mathbb{R}}
\newcommand{\IC}{\mathbb{C}}
\newcommand{\IF}{\mathbb{F}}
\newcommand{\IS}{\mathbb{S}}
\newcommand{\IV}{\mathbb{V}}
\newcommand{\II}{\mathbb{I}}
\newcommand{\IZ}{\mathbb{Z}}
\newcommand{\re}{{\rm Re}}
\newcommand{\im}{{\rm Im}}
\newcommand{\tr}{\mathop{\rm Tr}}
\newcommand{\ch}{{\rm ch}}
\newcommand{\rk}{{\rm rk}}
\newcommand{\ext}{{\rm Ext}}
\newcommand{\bi}{\begin{itemize}}
\newcommand{\ei}{\end{itemize}}
\newcommand{\beq}{\begin{equation}}
\newcommand{\eeq}{\end{equation}}

\newcommand{\CN}{{\cal N}}
\newcommand{\y}{{\mathbf y}}
\newcommand{\z}{{\mathbf z}}
\newcommand{\C}{\mathbb C}\newcommand{\R}{\mathbb R}
\newcommand{\CA}{\mathbb A}
\newcommand{\CP}{\mathbb P}
\newcommand{\cP}{\mathcal P}
\newcommand{\tmat}[1]{{\tiny \left(\begin{matrix} #1 \end{matrix}\right)}}
\newcommand{\mat}[1]{\left(\begin{matrix} #1 \end{matrix}\right)}
\newcommand{\diff}[2]{\frac{\partial #1}{\partial #2}}
\newcommand{\gen}[1]{\langle #1 \rangle}

\newtheorem{theorem}{\bf THEOREM}
\newtheorem{proposition}{\bf PROPOSITION}
\newtheorem{observation}{\bf OBSERVATION}
\newtheorem{statement}{\bf STATEMENT}

\def\theequation{\thesection.\arabic{equation}}
\newcommand{\setall}{
	\setcounter{equation}{0}
}
\renewcommand{\thefootnote}{\fnsymbol{footnote}}

\begin{titlepage}
\vfill
\begin{flushright}
{\tt\normalsize KIAS-P23039}\\

\end{flushright}
\vfill
\begin{center}
{\Large\bf $\mathbb{Z}_N$ Duality and Parafermions Revisited}

\vskip 1.5cm

Zhihao Duan, Qiang Jia and Sungjay Lee
\vskip 5mm

{\it School of Physics,
Korea Institute for Advanced Study, Seoul 02455, Korea}

\end{center}
\vfill

\begin{abstract}
Given a two-dimensional bosonic theory with a non-anomalous $\mathbb{Z}_2$ symmetry, the orbifolding and fermionization can be understood holographically  using three-dimensional BF theory with level $2$. From a Hamiltonian perspective, the information of dualities is encoded in a topological boundary state which is defined as an eigenstate of certain Wilson loop operators (anyons) in the bulk. We generalize this story to two-dimensional theories with non-anomalous $\mathbb{Z}_N$ symmetry, focusing on parafermionization. We find the generic operators defining different topological boundary states including orbifolding and parafermionization with $\mathbb{Z}_N$ or subgroups of $\mathbb{Z}_N$, and discuss their algebraic properties as well as the $\mathbb{Z}_N$ duality web.

\end{abstract}

\vfill
\end{titlepage}

\tableofcontents%\newpage
%\renewcommand{\thefootnote}{\#\arabic{footnote}}
%\setcounter{footnote}{0}
%\vskip 2cm

\section{Introduction and Conclusion}

Symmetries have consistently played an important role in the study of 
quantum field theories (QFTs). They often serve as guiding principles for
theoretical explorations. In recent years our understanding of symmetries 
has been further developed, advocating that ordinary (0-form) symmetries in QFTs, either continuous or discrete, can be best described by 
certain topological defects of co-dimension one. Such topological defects 
are referred to as symmetry operators. %Relying on this modern view, 

This perspective has catalyzed recent progress on generalizing 
the notion of symmetry in diverse directions. They encompass 
higher-form or higher-group symmetries \cite{Gaiotto:2014kfa,Albertini:2020mdx,Morrison:2020ool,Cordova:2018cvg,Cordova:2020tij,Bhardwaj:2020phs,Tian:2021cif,DelZotto:2022joo,DelZotto:2022fnw,Wang:2023iqt}, non-invertible symmetries \cite{Bhardwaj:2017xup,Chang:2018iay,Thorngren:2019iar,Komargodski:2020mxz,Thorngren:2021yso,Choi:2021kmx,Kaidi:2021xfk,Choi:2022zal,Cordova:2022ieu,Choi:2022jqy,Chang:2022hud,Bashmakov:2022uek}, subsystem symmetries \cite{Seiberg:2019vrp,Seiberg:2020bhn,Seiberg:2020wsg,Seiberg:2020cxy,Yamaguchi:2021xeq,Stahl:2021sgi,Gorantla:2022eem,Katsura:2022xkg,Gorantla:2022ssr,Yamaguchi:2022apr,Cao:2022lig,Cao:2023doz}, etc. A comprehensive aggregation of the vast literature on this 
subject can be found in \cite{Cordova:2022ruw}. See also 
\cite{Schafer-Nameki:2023jdn,Brennan:2023mmt,Bhardwaj:2023kri,Luo:2023ive}
for accessible reviews on these topics. 

Given a (generalized) symmetry, there exist various symmetry operations 
such as gauging and stacking invertible phases onto a given system. 
Many of them are blind to the details of dynamics but 
strongly tied with symmetry itself, and thus exhibit universality. 
The Symmetry Topological Field Theory (TFT) construction 
was recently proposed to provide an unified picture to 
study such symmetry operations. It is a framework where 
a given $d$-dimensional system of our interest 
is extended to $(d+1)$-dimensional slab with two boundaries. 
All dynamical information is encoded in one boundary 
while all symmetry manipulations take place in the other boundary. 
The bulk physics is governed by a topological field theory, 
and thus one can freely shrink the system and recover the original
theory. The Symmetry TFT has shed a new light on symmetry operations 
in a number of recent studies \cite{Gaiotto:2020iye,Apruzzi:2021nmk,Lin:2022dhv,Kaidi:2022cpf,vanBeest:2022fss,Kaidi:2023maf,Bhardwaj:2023ayw,Bartsch:2023wvv,Chen:2023qnv}. Note also that a closely related and mathematically rigorous framework is put forward in \cite{Freed:2022qnc}.

Our present work is partly motivated by the quest for even more insights from the
Symmetry TFT construction.
Specifically, we focus on a 2d quantum field theory $\mathcal{T}$ with a non-anomalous discrete 0-form symmetry $G$. Gauging the group $G$ results in an orbifold theory $\mathcal{T}/G$ with an emergent quantum symmetry $\tilde{G}$. Intriguingly, 
further gauging $\widetilde{G}$ reverts the system back to the theory $\mathcal{T}$ 
we begin by. In addition, we can also stack a (spin) TQFT phase on the system prior to 
the gauging. Altogether they leads to a rich zoo of interrelated theories. 

An illustrative but prominent example is the two-dimensional Ising conformal 
field theory (CFT). The Ising CFT has the non-anomalous 
$\mathbb{Z}_2$ symmetry. The Kramers-Wannier duality and 
the Jordan-Wigner transformation are two well-known symmetry 
manipulations giving rise to dual theories. The former maps the Ising 
CFT into itself with the order/disorder parameter interchanged. For the latter, 
we first couple the Ising CFT with the Kitaev Majorana chain in the topological phase,
and gauge the diagonal $\mathbb{Z}_2$ symmetry of the coupled system 
\cite{Gaiotto:2015zta}. The Jordan-Wigner transformation then fermionizes the Ising CFT to a system of free Majorana fermion where the emergent quantum $\mathbb{Z}_2$ symmetry is simply $(-1)^F$ with $F$ the fermion number. Note that further gauging the emergent $\mathbb{Z}_2$ symmetry maps the fermionic theory back to the Ising CFT. 
Built from all $\mathbb{Z}_2$ symmetry operations, one thus obtains 
two-dimensional duality web \cite{Karch:2019lnn,Gaiotto:2020iye,Hsieh:2020uwb,Kulp:2020iet} that 
interconnects various bosonic and fermionic theories. 

\begin{figure}[t!]
    \centering
    \includegraphics[width=12cm]{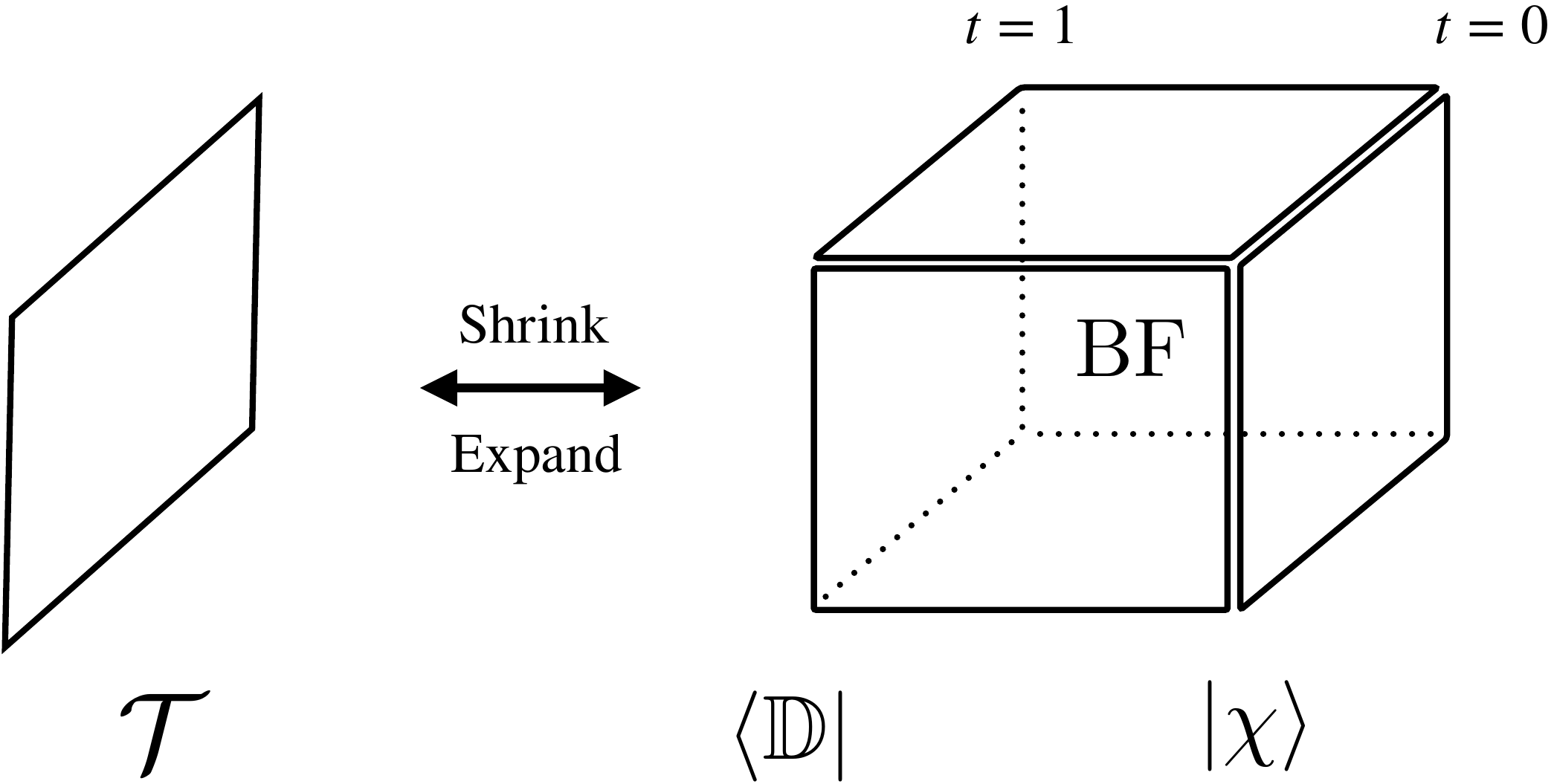}
    \caption{The Symmetry TFT construction. For the $\mathbb{Z}_N$ symmetry considered in this paper, we can choose the bulk TQFT to be the BF theory.}
    \label{fig:sandwich}
\end{figure}

Let us delve into the Symmetry TFT framework that offers 
an unified picture to understand various operations of 
gauging a discrete symmetry. 
The Symmetry TFT is a three-dimensional topological field theory 
placed on a slab $[0,1] \times \Sigma_g$ where $\Sigma_g$ is a genus-$g$ Riemann surface.
The bulk TFT for $\mathbb{Z}_2$ is the BF model with level two. 
When the interval $[0,1]$ is identified as the time direction, the initial 
and final states at $t=0$ and $t=1$ can be specified by the boundary conditions.  
For the boundary at $t= 0$ on the right hand side, 
we impose the so-called dynamical boundary condition that introduces an initial state $|\chi\rangle$. It is a state in the Hilbert space of the BF model on $\Sigma_g$ that 
accommodates all the dynamical information of the Ising model. On the other hand, 
we impose a topological boundary condition at $t=1$ so that a final state $\langle \mathbb{D}|$ only encodes the symmetry information of the 2d theory. 
As will be discussed in the main context, topological boundary states can 
be characterized by the type of anyon defects being diagonalized. Moreover,
they can be created by the condensation of the corresponding anyon. 
One can argue that the path-integral of the BF model on the slab then computes $\langle \mathbb{D}|\chi\rangle$ that agrees with the 
the partition function of a two-dimensional theory on $\Sigma_g$. 
A different choice of topological boundary state $\langle \mathbb{D}|$
results in the partition function of a different 2d theories 
involved in the duality web. Schematically the whole idea can be summarized
in Figure \ref{fig:sandwich}.

In the present work, we explore various duality maps between two-dimensional theories 
with non-anomalous $\mathbb{Z}_N$ symmetry. In the web of $\mathbb{Z}_N$ duality, 
one expects that fermionic theories should be generalized to the so-called 
$\mathbb{Z}_N$ parafermionic theories. The parafermionic theory refers to a theory that
contains operators obeying fractional statistics. 
Recall that, in order to define a 
fermionic theory on a Riemann surface, we need to specify the spin structure. 
Similarly, a parafermion theory depends on a choice of $\mathbb{Z}_N$ paraspin structure. 
As a natural generalization of emergent Majorana fermion from the critical Ising model,
the parafermion emerged from the the critical $\mathbb{Z}_N$-clock model 
\cite{FRADKIN19801, FATEEV1987644,GEPNER1987423}. 
Recently, it is revisited or studied in  \cite{Yao:2020dqx,Thorngren:2021yso,Burbano:2021loy,Haghighat:2023sax}, and a $\mathbb{Z}_N$ duality web was proposed in \cite{Thorngren:2021yso}. 

From the perspective of the Symmetry TFT, the bulk theory 
is naturally generalized to the BF model with level $N$. 
We then question if the Symmetry TFT picture can illuminate novel 
insights on the $\mathbb{Z}_N$ duality web. 
Our primary focus is to understand the connections between 
bulk anyonic operators, topological boundary states, and the 
$\mathbb{Z}_N$ orbifolds/parafermionic theories.
Naively one may expect the topological boundary states in $\mathbb{Z}_N$ are still eigenstates of the same type of anyonic operators as in the $\mathbb{Z}_2$ case. 
However, one encounters challenges in generalizing the fermionic topological boundary state.  This is because the corresponding anyon operators no longer commute with each other for $N \geq 3$. Instead, we propose a generic set of maximally commuting operators that exactly give rise to parafermionic topological boundary states. We then consider the 2d surface to be a torus, and identify all the topological boundary states in the $\mathbb{Z}_N$ duality web. We also discuss the intricate modular transformation of torus partition functions from the bulk perspective and demonstrate it for the duality web of $SU(2)_N/U(1)_{2N}$ coset CFTs, generalizing previous results in \cite{Yao:2020dqx}. When $N$ is not a prime number, one can partially gauge a subgroup
of $\mathbb{Z}_N$. We observe from the Symmetry TFT picture that the partial gauging for
either orbifold or parafermionic theory leads to a mixed 't Hooft 
anomaly in general. The anomalous phase we observed is consistent with the results 
of  \cite{Bhardwaj:2017xup}.

As for some possible future directions, first it would be nice to understand better the nature of the parafermionic operators and consequently the anyon condensation on the parafermionic boundary. Also, although in this paper we only analyze the parafermionization on a torus, it is tempting to generalize it to a higher genus Riemann surface. At present, this is still unclear partially due to a lack of understanding of the paraspin structure beyond the torus. Moreover, we can consider a higher dimensional setup, where a 4d gauge theory with 1-form symmetry is expanded to a 5d Symmetry TFT. It turns out that one can define surface operators that obey very similar algebraic relations as the parafermionic loop operators, and its implications is under investigation \cite{DJL24}.

This paper is organized as follows. In Section \ref{sec:2}, we give an overview on the $\mathbb{Z}_2$ duality web from both the 2d boundary and 3d Symmetry TFT perspective \cite{Karch:2019lnn,Hsieh:2020uwb,yujitasi}. In Section \ref{sec:3}, we generalize the discussion to the $\mathbb{Z}_N$ duality web \cite{Thorngren:2021yso} and study the interplay between the TFT bulk and 2d bosonic or parafermionic boundary theories. In Section \ref{sec:4}, we apply the general results obtained in the previous section to the $SU(2)_N/U(1)_{2N}$ coset CFTs as a concrete class of examples. Finally, Appendix \ref{App:operators} contains more details on the algebra of bulk anyonic operators. 

{\it Note Added:} While this work was in progress, we were informed of \cite{Chen:2023jht} where the authors study parafermions from the para-fusion fusion category perspective.

\section{A review on the web of 2d dualities: $\mathbb{Z}_2$ case}\label{sec:2}

Let us consider a two-dimensional bosonic theory with non-anomalous 
${\mathbb Z}_2$ symmetry. 
We review in this section two different procedures of  ${\mathbb Z}_2$ gauging, 
the Kramers-Wannier duality and the Jordan-Wigner transformation, that generate
the 2d duality web. We then discuss the BF model with level two in three dimensions, a.k.a. toric code 
that provides a bulk point of view of the two-dimensional dualities.

\subsection{$\mathbb{Z}_2$ orbifold and fermionization}

Let us start with a bosonic theory ${\cal B}$ with non-anomalous 
$\mathbb{Z}_2$ symmetry. Gauging the discrete $\mathbb{Z}_2$ symmetry,
one can obtain the orbifold ${\cal B}/\mathbb{Z}_2$. Note that 
the orbifold has an emergent quantum $\widetilde{\mathbb{Z}}_2$ symmetry. 
The Kramers-Wannier duality relates the partition function 
of ${\cal B}$ to that of ${\cal B}/\mathbb{Z}_2$ on a Riemann 
surface of genus $g$, $\Sigma_g$: 
\begin{equation}\label{Z2-Orbifolding}
	Z_{\widetilde{\mathcal{B}}}[\tilde{a}] = \frac{1}{2^g} \sum_{a\in H^1(\Sigma_g,\mathbb{Z}_2)}  \omega^{\langle \tilde{a} , a \rangle} Z_{\mathcal{B}}[a],
\end{equation}
where $\omega=e^{i\pi}$ is the second root of unity. 
Here $Z_{\cal B}[a]$ refers to the partition function of the theory ${\cal B}$ 
coupled to the $\mathbb{Z}_2$ background gauge field with holonomy $a$\footnote{We use $a$ for either holonomy or one-form.} while 
$Z_{\widetilde{\cal B}} [{\tilde a}]$ to the partition function of $\widetilde{\cal B}$
with $\widetilde{\mathbb{Z}}_2$ holonomy $\tilde a$. $\langle a, \tilde a \rangle$ 
is the anti-symmetric intersection pairing, which is defined as   
$\langle a, \tilde a \rangle \equiv \int a \cup \tilde a$.

In order to construct the fermionic theory ${\cal F}$ dual to 
${\cal B}$, we need to remind ourselves a few simple facts 
about a Majorana fermion. 
To define the theory of a free (massive) Majorana fermion on $\Sigma_g$,  
one has to specify the spin structure $s$ of the Riemann surface. 
The convention is that when $s_I=0$ the fermion satisfies Neveu-Schwarz (NS) boundary condition along $\Gamma_I$ 
and when $s_I=1$ it satisfies Ramond (R) boundary condition instead.
One can argue that the partition function of the gapped Majorana fermion 
in the topological phase, often referred to as the symmetry-protected-topological (SPT) phase of the Kitaev Majorana chain 
in modern language, is given by
\begin{align}
    Z_\text{Maj}^\text{top}[s] = (-1)^\text{Arf(s)},  
\end{align}
where $s$ is a choice of spin structure and 
the Arf$(s)$ is the Arf invariant. Upon the choice of 
symplectic basis $\Gamma_I \subset H_1(X,\mathbb{Z}_2)\ (I=1,2,\cdots,2g)$ with intersection pairing $\Gamma_I \cdot \Gamma_J = -\Gamma_J \cdot \Gamma_I = \epsilon_{IJ}$ with
\begin{equation}
	\epsilon_{IJ} = \left( \begin{array}{cc}
	0&I_{g\times g}\\
	-I_{g\times g}&0
	\end{array}\right).
\end{equation}
the Arf invariant is defined as 
\begin{equation}
	\textrm{Arf}(s) = \sum_{i=1}^g s_i s_{i+g} \quad \textrm{mod}\ 2.
\end{equation}
It is shown in \cite{ASENS_1971_4_4_1_47_0} that the Arf invariant coincides with the mod 2 index of the chiral Dirac operator on $\Sigma_g$. In other words, it counts the number of fermionic zero modes modulo $2$. 
It is then obvious that the Arf invariant becomes non-trivial 
for $2^{g-1}(2^g-1)$ spin structures of $\Sigma_g$ and trivial 
upon other $2^{g-1}(2^g+1)$ choices. For instance, let us consider $\Sigma_g = T^2$. 
One can argue that the Arf invariant is 
\begin{equation}
	\textrm{Arf}(s) = \left\{ \begin{array}{lc}
	1&(\textrm{R},\textrm{R})\\0&(\textrm{NS},\textrm{R}),(\textrm{R},\textrm{NS}),(\textrm{NS},\textrm{NS})
	\end{array}\right.
\end{equation}

The Kitaev Majorana chain plays a crucial role to perform  
the Jordan-Wigner transformation, a.k.a. fermionization, which maps a given ${\cal B}$ 
to its fermionic dual ${\cal F}$. To see this, let us first stack 
the Kitaev Majorana chain on the bosonic theory ${\cal B}$. 
Each of the two theories has a $\mathbb{Z}_2$ symmetry. By gauging 
the diagonal $\mathbb{Z}_2$ symmetry, one can then construct 
the fermionic theory ${\cal F}$ dual to ${\cal B}$, 
\begin{align}
    {\cal F} = \frac{{\cal B}\times \text{Kitaev}}{\mathbb{Z}_2}\ .
\end{align}
At the level of partition functions, the Jordan-Wigner transformation 
can be described as follows 
\begin{equation}
	Z_{\mathcal{F}}[s] = \frac{1}{2^g} \sum_{a\in H^1(\Sigma_g,\mathbb{Z}_2)}
    \omega^{\textrm{Arf}(s+a)} Z_{\mathcal{B}}[a] . 
\end{equation}

One can also construct another fermionic theory $\widetilde{\cal F}$ 
starting from the orbifold $\widetilde{\cal B}$, closely related to ${\cal F}$. 
To do so, all we need is to gauge the diagonal subgroup of 
the quantum $\widetilde{\mathbb{Z}}_2$ symmetry of $\widetilde{\cal B}$ and 
the fermion parity symmetry of the Kitaev Majorana chain. To be more explicit, 
the partition function of $\widetilde{\cal F}$ on $\Sigma_g$ with spin structure $\tilde{s}$ 
can be written as  
\begin{equation}
	Z_{\widetilde{\mathcal{F}}}[\tilde{s}] = \frac{1}{2^g} \sum_{\tilde{a}\in H^1(\Sigma_g,\mathbb{Z}_2)} \omega^{\textrm{Arf}(\tilde{s}+\tilde{a})} Z_{\widetilde{\mathcal{B}}}[\tilde{a}] .
\end{equation}
Using \eqref{Z2-Orbifolding} together with an identity of the Arf invariant, 
\begin{equation}\label{eq:Arfrule}
	\textrm{Arf}(s+b)= \textrm{Arf}(s) + \textrm{Arf}(b) + \langle s, b \rangle \quad \textrm{mod}\ 2,
\end{equation}
one can see that $\widetilde{\mathcal{F}}$ can also be obtained by stacking the SPT phase of Kitaev Majorana chain on ${\cal F}$, 
\begin{equation}\label{stack}
	Z_{\widetilde{\mathcal{F}}}[s] = \omega^{\textrm{Arf}(s)} Z_{\mathcal{F}}[s].
\end{equation}

We can summarize the duality web of ${\cal B}$, $\widetilde{\cal B}$, 
${\cal F}$, and $\widetilde{\cal F}$ in Diagram \ref{diag:fermion} \cite{Karch:2019lnn,Hsieh:2020uwb,yujitasi}. 
\begin{table}[t]
\begin{center}
        \begin{tikzpicture}[node distance = 6cm, thick]% 
        \node (B) {$\mathcal{B}$};
        \node (Bdual) [below= 3cm of B] {$\widetilde{\mathcal{B}} = \mathcal{B}/\mathbb{Z}_2$};
        \node (F) [right=of B] {$\mathcal{F}$};
        \node (Fdual) [below= 3cm of F] {$\widetilde{\mathcal{F}} = \mathcal{F}\otimes \omega^{\textrm{Arf}(s)}$};
        \draw[<->] (B) -- node [midway,above] {Fermionization} node [midway,below] {Bosonization} (F);
        \draw[<->] (B) -- node [sloped,below]{Orbifolding} (Bdual);
        \draw[<->] (Bdual) -- node [midway,above] {Fermionization} node [midway,below] {Bosonization} (Fdual);
        \draw[<->] (F) -- node [sloped,above]{Stacking} (Fdual);
    \end{tikzpicture}
\end{center}
\caption{The web of $\bZ_2$ dualities.}
\label{diag:fermion}
\end{table}

%%%%%%%%%%%%%%%%%%%%%%%%%%%%%%%%%%%%%%%%%%%%%%%%%%%%%%%%%%%%%%%%%%%%%%%%%%%%%%%%%%%%%%%%%%%%
%%%%%%%%%%%%%%%%%%%%%%%%%%%%%%%%%%%%%%%%%%%%%%%%%%%%%%%%%%%%%%%%%%%%%%%%%%%%%%%%%%%%%%%%%%%%

\subsection{BF theory and topological boundary}\label{sec:Z2boundary}

As mentioned in Introduction, the Symmetry TFT provides 
a unified and coherent picture to understand the above 
web of two-dimensional dualities. Since the theories 
involved in the duality web have a $\mathbb{Z}_2$ symmetry, 
the Symmetry TFT of our interest is the BF model 
with level two, 
\begin{align}\label{BFmodel}
    S_{BF} = \frac{N}{2\pi} \int \tilde{A} \wedge d A,
\end{align}
with $N=2$. It is a three-dimensional $\mathbb{Z}_2$ gauge 
theory, and is often called the toric code in the condensed 
matter literature. 

For later convenience, let us first place the BF model 
on a spatial torus, $\Sigma_g=T^2$. Since 
the given theory is topological, the Hilbert space ${\cal H}_{T^2}$ can 
be obtained by quantizing the space of (classical) 
vacuum states. The vacuum classical field configurations 
are flat connections and can be described as their holonomies 
around cycles in $\Sigma_g=T^2$. To be concrete, one can 
represent the flat connections on $\Sigma_g=T^2$ as constant 
gauge fields 
\begin{align}
    A_i =a_i /N  \ , \quad \tilde A_i = \tilde a_i/N \ . 
\end{align}
Here both $a_i$ and $\tilde a_i$ ($i=1,2$) are 
normalized so that they are defined modulo $N$ shifts, i.e.,
$a_i \simeq a_i + N$ and $\tilde a_i \simeq \tilde a_i +N$. Let 
the holomonies vary slowly over time, and plug them back into 
the action. The low-energy effective action becomes  
\begin{align}
    S_{BF}^\text{eff} = \frac{2\pi}{N} 
    \int dt \left( 
    \tilde a_2 \frac{d}{dt} a_1 - 
    \tilde a_1 \frac{d}{dt} a_2 \right)\ . 
\end{align}

Upon the canonical quantization, 
\begin{align}\label{quantizationcan}
    \big[ \hat{a}_i , \frac{2\pi}{N} \hat{\tilde{a}}_j \big] & =   i \e_{ij},
\end{align}
with $\e_{12}=-\e_{21}= + 1$, we have two canonical bases of the Hilbert space.
One of them is a set of `position' eigenstates,  
\begin{align}\label{basis01}
    \text{exp}\Big[  \frac{2\pi i}{N} \hat{a}_i \Big] | a \rangle  & = \omega^{a_i} |a \rangle,  
    \nonumber \\ 
    \text{exp}\Big[ \frac{2\pi i}{N} \hat{\tilde{a}}_j \Big] | a \rangle & = | a - \gamma_j \rangle  , 
\end{align}
and the other is  a set of `momentum' eigenstates, 
\begin{align}\label{basis02}
    \text{exp}\Big[  \frac{2\pi i}{N} \hat{a}_i \Big] |{\tilde a}  \rangle & = | {\tilde a}  - \gamma_i \rangle
    \nonumber \\ 
    \text{exp}\Big[ \frac{2\pi i}{N} \hat{\tilde{a}}_j \Big] | {\tilde a} \rangle  & =  \omega^{{\tilde a}_j} |{\tilde a} \rangle,   
\end{align}
where $\omega=e^{2\pi i /N}$ is the $N$-th root of unity and $\gamma_i$ are unit vectors with non-vanishing components $(\gamma_i)_j=-\e_{ij}$.  
One can also show that
\begin{align} \label{Fourier01}
    \langle a | {\tilde a} \rangle = \frac1N \text{exp}\Big[  \frac{2\pi i }{N} \e_{ij} a_i {\tilde a}_j \Big].
\end{align}
Since $a_i$ and $\tilde a_i$ are periodic,  both position and momentum values are quantized,  
\begin{align}\label{quantization}
    a_ i & = 0,1,.., N-1\ ,  \nonumber \\   
    \tilde a_i & = 0 ,1, .., N-1 \ . 
\end{align}
It is then clear from \eqref{quantization} that the operators 
in \eqref{basis01} and \eqref{basis02}, for any $i$ and $j$, satisfy the relations below 
\begin{align}\label{relationont2}
    \left(  \text{exp}\Big[  \frac{2\pi i}{N} \hat{a}_i \Big] \right)^N = 
    \left(  \text{exp}\Big[  \frac{2\pi i}{N} \hat{\tilde{a}}_j \Big] \right)^N = {\bf 1}\ .
\end{align}
To express \eqref{basis01} and \eqref{basis02}  in terms of field variables, 
we first note that 
\begin{align}
    \text{exp}\Big[ i \oint_{\Gamma_i} A \Big]  & =  \text{exp}\Big[  \frac{2 \pi i}{N} \langle \gamma_i , \hat{a} \rangle \Big]\ , 
    \nonumber \\
    \text{exp}\Big[ i \oint_{\Gamma_i} \tilde{A} \Big]  & =  \text{exp}\Big[  \frac{2 \pi i}{N} \langle \gamma_i ,  \hat{\tilde{a}} \rangle \Big]\ ,    
\end{align}
where $\Gamma_i$ refers to the one-cycles along $x^i$ and 
the symplectic bilinear form is defined as $\langle a, b \rangle = \e_{ij} a_i b_j$.
The Wilson loops $U[\Gamma] = e^{i\oint_\Gamma A}$ and ${\tilde U}[\Gamma] = e^{i\oint_\Gamma {\tilde A}}$ 
are then diagonalized in which $\{ |a\rangle \}$ and $\{ | \tilde a \rangle \}$ 
are chosen as a basis of the Hilbert space of the BF model on $T^2$, respectively. 
From \eqref{quantizationcan}, one can show the loop operators satisfy the commutation relation below, 
\begin{align}\label{algebra01}
    U[\Gamma_i] \widetilde{U}[\Gamma_j] =  \omega^{-\langle \gamma_i, \gamma_j \rangle }   \widetilde{U}[\Gamma_j] U[\Gamma_i]\ .
\end{align}

Let us then place the BF model on a Riemann surface of arbitrary genus $g$, $\Sigma_g$. 
The observables in the theory are the Wilson loops, 
\begin{equation}
    L_{e,m} [\Gamma] = U^e[\Gamma] \widetilde{U}^m[\Gamma]
\end{equation}
with 
\begin{equation}
     U[\Gamma] = \exp \Big[i \oint_{\Gamma} A\Big],\quad  
     \widetilde{U}[\Gamma]=\exp \Big[i \oint_{\Gamma} \tilde{A} \Big]\ . 
\end{equation}
One can argue that the above loop carries the statistical spin 
\begin{align}
    s = \frac{e m }{N} \quad \text{mod} ~ 1. 
\end{align}
The operator relation \eqref{relationont2} on two-torus can be generalized to 
\begin{align}
    U^N[\Gamma] =  \widetilde{U}^N[\Gamma] = {\bf 1}
\end{align}
for any $\Gamma$ on $\Sigma_g$. This is consistent with the fact that those operators do not induce any 
holonomies and carry no spin. Thus, one should identify the loops $L_{(e,m)}$ labeled by $(e,m)$, $(e+N,m)$, and 
$(e,m+N)$. In other words, two charges $e$ and $m$ are $\mathbb{Z}_N$-valued.  
The loop operators satisfy the so-called quantum torus algebra on $\Sigma_g$
\begin{equation}\label{BF-algebra-relation-1}
	L_{e,m} [\Gamma]L_{e',m'} [\Gamma'] = \omega^{-(em'+me') \langle \gamma , \gamma' \rangle } L_{e',m'} [\Gamma'] L_{e,m} [\Gamma],
\end{equation}
and
\begin{equation}\label{BF-algebra-relation-2}
	L_{e,m}[\Gamma] L_{e,m}[\Gamma']=\omega^{-em \langle\gamma,\gamma'\rangle}L_{e,m}[\Gamma+\Gamma'],
\end{equation}
which essentially reduce to \eqref{algebra01} when $\Sigma_g = T^2$.
As an obvious generalization of $\gamma_i$ in \eqref{algebra01}, $\gamma$ and $\gamma'$ are the Poincar\'{e} dual of one-cycles $\Gamma$ and $\Gamma'$
on the Riemann surface $\Sigma_g$.  

\subsection{$\bZ_2$ duality web from the Symmetry TFT}

In this section, we revisit the aforementioned two-dimensional duality web to demonstrate how useful 
the Symmetry TFT picture is. To do so, we set $N=2$ for the BF model defined on 
a slab $[0,1] \times \Sigma_g$. The interval $[0,1]$ represents the time direction. One then 
propagates a state in the Hilbert space ${\cal H}_{\Sigma_g}$ from $t=0$ to $t=1$. The time-evolution is particularly 
simple because the BF model has a vanishing Hamiltonian.

To describe initial and final states at $t=0$ and $t=1$, 
we introduce a canonical basis of the Hilbert space of the BF model on $\Sigma_g$ where 
either $U[\Gamma]$ or $\widetilde{U}[\Gamma]$ are diagonalized: 
\begin{itemize}
    \item ($\mathcal{A}$-basis) Eigenstates of $U[\Gamma]$
        \begin{equation}\label{eq:Z2Ebasis}
        \left\{\begin{array}{l}
			U[\Gamma] |a\rangle = \omega^{\langle \gamma , a\rangle} |a\rangle\\
			\widetilde{U}[\Gamma] |a\rangle = |a - \gamma \rangle
            \end{array}\right.
        \end{equation}
    \item ($\widetilde{\mathcal{A}}$-basis) Eigenstates of $\widetilde{U}[\Gamma]$
        \begin{equation}\left\{\begin{array}{l}
			\widetilde{U}[\Gamma] |\tilde{a}\rangle = \omega^{\langle\gamma , \tilde{a}\rangle} |\tilde{a}\rangle\\
			U[\Gamma] |\tilde{a}\rangle = |\tilde{a} - \gamma \rangle
            \end{array}\right.
        \end{equation}
\end{itemize}
where $\omega=\exp(i \pi)$ and $a,\tilde{a}$ are $\mathbb{Z}_2$-valued 
holonomies on the Riemann surface $\Sigma_g$.
As in \eqref{Fourier01}, the two bases are related by a discrete Fourier transformation,
\begin{equation}\label{eq:atilde}
    |\tilde{a}\rangle = \frac{1}{2^g} \sum_{a} \omega^{\langle a,\tilde{a}\rangle} |a\rangle\ .
\end{equation}

Specifying two boundary conditions at $t=0$ and $t=1$, the path-integral computes an inner product between corresponding boundary states.
In what follows, an initial state at $t=0$ is always fixed by the so-called `dynamical' boundary state $|\chi \rangle$. 
One can construct the state by coupling the topological BF model to a given two-dimensional theory ${\cal B}$ 
on the boundary $\Sigma_g$ at $t=0$.
Let a final state at $t=1$ be an eigenstate of the Wilson loop $U$, $|a\rangle$. 
Then, the path-integral of the BF model on the slab gives
\begin{align}
    Z = \langle a | e^{i H t} | \chi \rangle = \langle a | \chi   \rangle\ ,
\end{align}
which has to agree with the torus-partition function of ${\cal B}$ in the 
presence of the $\mathbb{Z}_2$ background $a$, namely, $Z_{\mathcal{B}}[a]$. 
It implies that the dynamical state can be expressed as 
\begin{equation}
    |\chi\rangle = \sum_{a} Z_{\mathcal{B}}[a] |a\rangle\ .
\end{equation}
If we choose a different initial state, say $| \tilde a \rangle $, 
the path-integral on the slab  computes  
\begin{align}
    Z= \langle \tilde a | \chi \rangle   =  
    \frac{1}{2^g} \sum_a \omega^{ \langle \tilde a , a \rangle } Z_{\mathcal{B}}[a] , 
\end{align}
where we used \eqref{eq:atilde} for the last equality. 
It is nothing but the partition function of the orbifold ${\cal \widetilde{B}}$
coupled to the ${\tilde a}$ holonomy. In other words, the $\mathbb{Z}_2$ 
gauging of ${\cal B}$ can be viewed from the Symmetry TFT simply as   
switching a final state from $| a \rangle$ to $| \tilde a \rangle$.

One can obtain the boundary states $|a \rangle$ and $|\tilde a \rangle$ 
by imposing the Dirichlet boundary condition for $A$ and $\tilde A$ at $t=1$, 
respectively. However, let us explain a different but convenient description 
of such states proposed in \cite{Witten:1988hf}. For simplicity, we focus on the case 
where $\Sigma_g=T^2$. 
A choice of one-cycle of $T^2$ that becomes non-contractible 
inside the solid torus determines how to identify $T^2$ 
as the boundary of the solid torus. We place a Wilson loop, 
either $U$ or $\widetilde{U}$, inside the solid torus. 
Performing the path-integral then creates a boundary state 
$|a\rangle $ or $| \tilde a \rangle$ on $T^2$.  
For instance, to obtain a state $|a\rangle$ with $a=na_1 + m a_2$ on the boundary $T^2$, 
one requires that $n \Gamma_1 + m \Gamma_2$ (two cycles $\Gamma_I$ compose a basis of $H_1(T^2,\mathbb{Z}_N)$) is 
the non-contractible cycle and place the Wilson loop $U$ inside the solid torus.  
The states created in this manner are often referred to as topological 
boundary states in the literature.

What if we insert a different loop operator $U_F=L_{1,1}$ at the core of 
the solid torus? It does create a different set of topological boundary states on $T^2$, labelled by $|s \rangle$. 
One can see from \eqref{BF-algebra-relation-1} that the states $\{ |s \rangle\} $ can simultaneously diagonalize $U_F[\Gamma]=L_{1,1}[\Gamma]$  
with eigenvalues $\omega^{-Q_s(\gamma)}$ for all $\Gamma \in H_1(\Sigma_g, \mathbb{Z}_2)$,
\begin{align}
    U_F[\Gamma] |s\rangle = \omega^{-Q_s(\gamma)} |s\rangle. 
\end{align}
Due to \eqref{BF-algebra-relation-2}, the eigenvalues are constrained to satisfy a relation below 
\begin{align}
    \omega^{Q_s(\gamma+\gamma') } =  \omega^{Q_s(\gamma) + Q_s(\gamma') + \langle \gamma, \gamma' \rangle }. 
\end{align}
In other words, the function $Q_s(\gamma)$ has to be a quadratic 
refinement of the symmetric pairing $\langle \ast, \ast \rangle$\footnote{When the 1-cocycles are $\mathbb{Z}_2$ valued, the anti-symmetric pairing 
$\langle \ast , \ast \rangle$ becomes symmetric mod 2.}. 
Moreover, since $U_F(\Gamma)$ carries spin $1/2$, a `holonomy' $s$ should specify a spin structure 
of $\Sigma_g$. It was shown in \cite{Karch:2019lnn} that those two requirements 
are satisfied by expressing $Q_s(\gamma)$ in terms of the Arf invariant, %{\nb{ZH: do the two requirements "determine"?}}
\begin{align}\label{quadraticrefinement01}
    Q_s(\gamma) = \text{Arf}(s+\gamma) - \text{Arf}(s) \text{ mod } 2 \ .
\end{align}
It also implies that the topological boundary state $|s\rangle $ can be expressed as 
\begin{align}\label{fermionbasis}
    | s \rangle = \frac{1}{2^g} \sum_a \omega^{-\text{Arf}(s+a) } |a \rangle\ . 
\end{align}
To see this, let $U_F[\Gamma]$ act on $|s\rangle$:
\begin{align}
    U_F[\Gamma] |s \rangle & = \frac{1}{2^g} \sum_a  \omega^{-\text{Arf}(s+a) } \cdot U[\Gamma] \widetilde{U}[\Gamma] |a \rangle
    \nonumber \\ & = 
    \frac{1}{2^g} \sum_a  \omega^{-\text{Arf}(s+a + \gamma) + \langle \gamma, a \rangle } | a \rangle
    \nonumber \\ & = 
    \omega^{-\text{Arf}(\gamma+s) + \text{Arf}(s)} |s \rangle\ , 
\end{align}
where we use for the last equality an identity for the Arf invariant 
\begin{align}
    \text{Arf}( s + a + \gamma) = \text{Arf}( \gamma + s) + \text{Arf}(a + s ) + \text{Arf}(s) + \langle \gamma, a \rangle \text{ mod } 2\ . 
\end{align}

If we choose $|s \rangle$ as a final state, the path-integral on the slab computes
the transition amplitude, 
\begin{align}
    Z = \langle s | \chi \rangle = \frac{1}{2^g} \sum_{a} \omega^{\text{Arf}(s+a)} Z_{\cal B}[a] \ ,
\end{align}
which coincides with the partition function of the fermionic theory ${\cal F}$ dual to ${\cal B}$. Thus, one can say that the Jordan-Wigner transformation can be described as the change of a topological boundary state from $|a\rangle$ to $|s\rangle$. 

Analogous to \eqref{fermionbasis}, one may use $\{| \tilde a \rangle \}$ 
to define a state $\widetilde{|s\rangle}$ as   
\begin{align}\label{tildesN2}
     \widetilde{|s\rangle} = \frac{1}{2^g} \sum_a \omega^{-\text{Arf}(s+\tilde a) } |\tilde a \rangle\ .
\end{align}
However, the state $\widetilde{|s\rangle}$ is not a new topological boundary 
state but essentially the same as $|s\rangle$. 
More precisely, using an identity below, 
\begin{align}
    \frac{1}{2^g} \sum_a \omega^{\text{Arf}( s + a) + \text{Arf}(s) + \langle a, t \rangle } =    \omega^{\text{Arf}(t+s)}\ ,
\end{align}
one can show that 
\begin{align}
    \widetilde{|s\rangle} = \omega^{\text{Arf}(s)} | s \rangle\ .    
\end{align}
When we can choose $\widetilde{|s\rangle}$ as a final state, 
the path-integral results in the partition function of $\widetilde{\cal F}$,
\begin{align}
    \widetilde{\langle s |}   \chi \rangle = 
    \frac{1}{2^g}\sum_{\tilde a} \omega^{\text{Arf}(s+\tilde a)} 
    Z_{\tilde B}[\tilde a]\ . 
\end{align}

In summary, given a two-dimensional theory with a non-anomalous $\mathbb{Z}_2$ symmetry, one can associate it with a boundary state $|\chi\rangle$ belonging to the Hilbert space of the BF model with level two. The different theories in the duality web can be understood as projecting $|\chi\rangle$ onto different topological boundary states  $|a\rangle$, $|\tilde{a}\rangle$, and $|s\rangle$ that are eigenstates of the loop operators $U$, $\widetilde{U}$, and $U_{F}$ \cite{Gaiotto:2020iye,yujitasi}. As a final remark, in the condensed matter literature, these topological boundary states are related to the anyon condensation or fermion condensation \cite{kong2014anyon,Aasen:2017ubm}, and correspond to the electric, magnetic, and fermionic gapped boundaries of the toric code respectively.

%%%%%%%%%%%%%%%%%%%%%%%%%%%%%%%%%%%%%%%%%%%%%%%%%%%%%%%%%%%%%%%%%%%%%%%%%%%%%%%%%%%%%%%%%%%%%%%%%%%%%%%%%%%%%%%%%%%%%%%%%%%%%%%%%%%%%%
%%%%%%%%%%%%%%%%%%%%%%%%%%%%%%%%%%%%%%%%%%%%%%%%%%%%%%%%%%%%%%%%%%%%%%%%%%%%%%%%%%%%%%%%%%%%%%%%%%%%%%%%%%%%%%%%%%%%%%%%%%%%%%%%%%%%%%
%%%%%%%%%%%%%%%%%%%%%%%%%%%%%%%%%%%%%%%%%%%%%%%%%%%%%%%%%%%%%%%%%%%%%%%%%%%%%%%%%%%%%%%%%%%%%%%%%%%%%%%%%%%%%%%%%%%%%%%%%%%%%%%%%%%%%%

\section{Web of 2d dualities: $\mathbb{Z}_N$ case}\label{sec:3}

Based on the Symmetry TFTs, we present in this section a web of two-dimensional 
dualities for theories with $\mathbb{Z}_N$ symmetry. 
The Symmetry TFT of our interest is the BF model with $N>2$ \eqref{BFmodel} defined on the slab. 
Let us begin by a bosonic theory with non-anomalous $\mathbb{Z}_N$ symmetry.

The initial state at $t=0$ remains fixed by the dynamical 
boundary state $|\chi \rangle$. On the other hand, one 
also needs topological boundary states at $t=1$ to unveil the dualities. 
As explained earlier, the Hilbert space of the BF model on $\Sigma_g$ 
has two canonical bases, ${\cal A}$-basis and $\widetilde{\cal A}$-basis.
Both contain topological boundary states, $|a\rangle$ or
$|\tilde a\rangle$, and diagonalize the Wilson loop operators 
$U[\Gamma]$ or $\widetilde{U}[\Gamma]$. 
In terms of topological boundary states $|a\rangle$, 
the dynamical boundary state reads 
\begin{align}
    |\chi\rangle =  \sum_{a} Z_{\cal B}[a] |a \rangle\ ,
 \end{align}
where $Z_{\cal B}[a]$ is the partition function of ${\cal B}$ on $\Sigma_g$ 
coupled to the background $\mathbb{Z}_{N}$ holonomies $a$. 

The Symmetry TFT says that one can easily describe 
the  partition function of orbifold $\widetilde{\cal B}= {\cal B}/ \mathbb{Z}_N$ by replacing the final 
state from $|a\rangle$ to $|\tilde a\rangle$: 
To be concrete, it is given by 
\begin{align}
    Z_{\widetilde{\cal B}}[\tilde a]  = \langle \tilde a | \chi \rangle 
   = \frac{1}{N^g} \sum_a \omega^{\langle \tilde a , a \rangle} Z_{\cal B}[a], 
\end{align}
where $\omega=e^{2\pi i /N}$ is the $N$-th root of unity. 

As demonstrated earlier, the topological boundary states $\{ |s\rangle\}$ associated with 
the loop operators $U_F[\Gamma]=L_{1,1}[\Gamma]$ are essential to 
understand the Jordan-Wigner transformation from the Symmetry TFT point of view. 
However, one cannot define such states for $N>2$. This is partly because
the operators $U_F[\Gamma]$ no longer commute with each other when $N>2$. 
The commutation relation read from \eqref{BF-algebra-relation-1}   
\begin{align}
    U_F[\Gamma]  U_F[\Gamma'] = \omega^{-2 \langle \gamma, \gamma' \rangle}
    U_F[\Gamma'] U_F[\Gamma] \ , 
\end{align}
indeed has the phase factor that only becomes trivial for $N=2$. 
The obstruction motivates us to search for maximally commuting loop operators
other than $U[\Gamma]$ and $\widetilde{U}[\Gamma]$. If they exist, one can regard 
their eigenstates as topological boundary states that generalize fermionic states $\{ |s \rangle\}$ of $N=2$.

Let us first discuss how to characterize the loop operators $\{ L[\Gamma]\}$ that define topological boundary states. They are required to satisfy the relations below 
\begin{enumerate}

    \item The topological boundary states are by definition 
    eigenstates of the loop operators $L[\Gamma]$. Thus, 
    for any cycles $\Gamma$, $L[\Gamma]$ mutually commute 
    \begin{align}\label{condition01}
        \Big[ L[\Gamma] , L[\Gamma'] \Big]=0\ .
    \end{align}

    \item Since the BF model on $\Sigma_g$ has the Hilbert space of dimensions $N^{2g}$, 
    topological boundary states should be described by $\mathbb{Z}_N$-valued eigenvalues. It implies that any loop operators associated  with 
    topological boundary states must obey  
    \begin{align}\label{condition02}
        L^N[\Gamma] =1\ .     
    \end{align}

    \item The dimension of the Hilbert space also requires that 
    any product of two loop operators should not generate a new 
    independent loop operator, i.e., 
    \begin{align}\label{condition03}
        L[\Gamma] L[\Gamma'] = \epsilon(\gamma, \gamma') 
        L[\Gamma+\Gamma']\  . 
    \end{align}
    Here, \eqref{condition01} and \eqref{condition02} further constrain the factor $\epsilon(\gamma, \gamma')$ of
    \eqref{condition03} to satisfy the relations below  
    \begin{align}\label{condition03_01}
        \epsilon(\gamma,\gamma')  = \epsilon(\gamma', \gamma)\ , 
        \quad \epsilon^N(\gamma,\gamma') = 1\ .         
    \end{align}

\end{enumerate}
It is easy to show that the loop operators $U[\Gamma]$ and $\widetilde{U}[\Gamma]$ 
satisfy the above constraints with $\epsilon(\gamma,\gamma')=1$. We can also find 
other loop operators when the phase factor $\epsilon(\gamma,\gamma')$ is chosen as 
\begin{align}\label{condition03_02}
    \epsilon(\gamma, \gamma') = \omega^{k (\gamma,\gamma')}
\end{align}
where $k$ is an arbitrary integer coprime with $N$ and 
$(\ast,\ast)$ in the exponent is a symmetric pairing. We can argue that 
the corresponding topological boundary states, denoted by 
$\{| s;k \rangle \}$, can define the `parafermionization' of a given ${\cal B}$, natural 
generalization of the fermionization for $N>2$. 
We examine the properties the $\{ |s;k\rangle \}$, and present 
the explicit expression of those loop operators below.

\subsection{Parafermionization}

Let us denote by $U_{Pf;k}[\Gamma]$ the loop operators 
satisfying the constraints from \eqref{condition01} to \eqref{condition03} 
with \eqref{condition03_02}. Their explicit form will be presented shortly.

We first discuss various features of their eigenstates $\{| s;k \rangle \}$,
\begin{align}
    U_{Pf;k}[\Gamma] | s;k \rangle = \omega^{- Q_s(\gamma;k)} | s ;k \rangle 
\end{align}
relevant for generalizing the Jordan-Wigner transformation for $N>2$. 
As we can see later that the loop operators $U_{Pf;k}[\Gamma]$ carry the fractional spin $S=\pm k/N$, 
the $\mathbb{Z}_N$-valued holonomy $s$ could determine 
the $\mathbb{Z}_N$ version of the spin structures, known as paraspin structure. 
Due to the lack of complete understanding 
of the paraspin structure beyond the torus (see \cite{Runkel:2018feb} for another definition of paraspin structure), the Riemann surface 
$\Sigma_g$ is restricted to $T^2$ in what follows. 
The relation \eqref{condition03} implies that eigenvalues should 
satisfy 
\begin{align}\label{quadraticrefinement02}
    \omega^{Q_s(\gamma+\gamma';k)} = \omega^{Q_s(\gamma;k) + Q_s(\gamma';k) + k (\gamma,\gamma') } \ ,   
\end{align}
which suggests that $Q_s(\gamma;k)$ could be refereed to as 
a quadratic refinement of the symmetric pairing $(\ast,\ast)$. 
Our convention for the symmetric pairing is given by
\begin{align}
    ( a, b ) = a_1 b_2 + a_2 b_1\ .
\end{align}
One can show that a natural generalization of \eqref{quadraticrefinement01}
satisfies \eqref{quadraticrefinement02},
\begin{align}\label{quadrefinement03}
    Q_s(\gamma;k) = k \Big( \text{Arf}_N(s+\gamma) - \text{Arf}_N(s)  \Big) \text{ mod } N\ , 
\end{align}
where the $\mathbb{Z}_N$-valued Arf invariant \cite{Yao:2020dqx,Thorngren:2021yso} on a torus becomes
\begin{align}
    \text{Arf}_N(s) = s_1 s_2 \ . 
\end{align}

To be concrete, let us present an explicit expression of $U_{Pf;k}[\Gamma]$ below
\begin{align}\label{ZN-exotic-operators-UPFp}
    U_{Pf;k}[\Gamma] = 
    U[m \Gamma_1]^{-k} U[n\Gamma_2]^{k} \widetilde{U}[m\Gamma_1]  \widetilde{U}[n\Gamma_2]\ , 
\end{align}
where a given cycle $\Gamma$ is decomposed in terms of two generators $\Gamma_1$ and $\Gamma_2$ 
as  
\begin{align}
    \Gamma = m \Gamma_1 + n \Gamma_2\ .     
\end{align}
For more detailed algebraic relations of $U_{Pf;k}[\Gamma]$, we refer the reader to Appendix \ref{App:operators}. In particular, from \eqref{eq:commutation} we can read off the fractional spin of $U_{Pf;k}[\Gamma]$. When $N=2$ and $k=1$, one can see that \eqref{ZN-exotic-operators-UPFp} reduces to 
the $U_{F}[\Gamma]$. Note that the above construction depends on a choice of basis for $H_1(T^2,\mathbb{Z}_N)$ when $N>2$, 
reflecting the fact that the symmetric pairing with $N>2$ is not 
invariant under the $SL(2,\mathbb{Z})$ transformation. In other words, 
the corresponding topological boundary states $|s;k\rangle$ depends on 
the choice of basis, which eventually gives rise to
their intricate modular properties. We will elaborate on this aspect later. 

To represent the topological boundary states $|s;k\rangle$ in the basis 
$\{ |a \rangle$\}, we first note that 
the action of $U_{Pf;k}[\Gamma]$ on the state $|a\rangle$ becomes 
\begin{align}
    U_{Pf;k}[\Gamma] | a \rangle = \omega^{k(\gamma,a-\gamma)} | a -\gamma \rangle\ ,
\end{align}
where we used \eqref{eq:Z2Ebasis}. It implies that $ |s;k\rangle $ can be described as 
\begin{align}\label{eq:ZNs}
    |s; k\rangle = \frac{1}{N} \sum_a \omega^{- k \text{Arf}_N(s+a)} |a \rangle \ .
\end{align}
This is because $U_{Pf;k}[\Gamma]$  acts on the state as follows, 
\begin{align}
    U_{Pf;k}[\Gamma] |s; k\rangle & = 
    \frac1N \sum_a \omega^{- k \text{Arf}_N(s+a +\gamma) } \omega^{k(\gamma, a ) } |a \rangle\ ,
    \nonumber \\ & = \omega^{-k (\text{Arf}_N(s+\gamma) -\text{Arf}_N(s) ) } |s;k \rangle\ . 
\end{align}
For the last equality, we used an identity for the Arf$_N$ invariant  
\begin{align}
    \text{Arf}_N(s+a+\gamma) = \text{Arf}_N(s+a) + \text{Arf}_N(s+\gamma) - \text{Arf}_N(s) + (\gamma,a)  \text{ mod } N\ . 
\end{align}

When the final state at $t=1$ is given by $|s;k\rangle$ \eqref{eq:ZNs}, 
the BF model on the slab then provides  a $\mathbb{Z}_N$ generalization of 
the Jordan-Wigner transformation, 
\begin{align}\label{ZN-parafermionization}
    Z_{\cal{PF} }[s;k] \equiv \langle s;k | \chi \rangle = 
    \frac{1}{N} \sum_a \omega^{k \text{Arf}_N(s+a)} Z_{\cal B}[a] \ , 
\end{align}
where $Z_{\cal PF}[s;k]$ is the torus partition function of 
a dual `parafermionic' theory $\cal PF$ with the $\mathbb{Z}_N$ paraspin structure $s$. 
The parafermionic theory refers to a theory that contains operators obeying 
fractional statistics. The above result exactly agrees with the 
torus partition function proposed previously in \cite{Thorngren:2021yso}. In fact, 
\eqref{ZN-parafermionization} can be understood as the continuum version 
of the Fradkin-Kadanoff transformation \cite{FRADKIN19801} that maps the $\mathbb{Z}_N$ clock 
model to the $\mathbb{Z}_N$ parafermion.

Applying the parafermionization to the orbifold $\widetilde{\cal B}$ gives
\begin{align}
    Z_{\widetilde{\cal PF}}[s;k] = \frac{1}{N} \sum_{\tilde a} \omega^{k \text{Arf}_N(s+\tilde a)} Z_{\widetilde{\cal B}}[\tilde a],  
\end{align}
which can be described as the transition amplitude between the dynamical state $| \chi \rangle$ 
and a generalization of \eqref{tildesN2}, 
\begin{align}\label{eq:tildes01}
    \widetilde{| s;k \rangle} = \frac{1}{N} \sum_{\tilde a} \omega^{- k \text{Arf}_N(s+\tilde a)} | \tilde a\rangle\ . 
\end{align}
Rewriting \eqref{eq:tildes01} as 
\begin{align}
    \widetilde{| s;k \rangle} & = \left( \frac{1}{N} \right)^2 \sum_a \sum_{\tilde a}
    \omega^{- k \text{Arf}_N(s+\tilde a)} \omega^{k \langle a, \tilde a \rangle } | ka \rangle
    \nonumber \\ & =
    \frac1N \sum_{a}  \omega^{ - k\text{Arf}_N(s'+a ) - k \text{Arf}_N(s)} | ka \rangle\ , 
\end{align}
one can read off the relation between the state $\widetilde{| s; k\rangle}$ and the parafermion state  
\begin{align}
    \omega^{k \text{Arf}_N(s) } \widetilde{| s;k \rangle} =   | ks' ; \frac1k \rangle \ .
\end{align}
Here $s'=(-s_1,s_2)$ and $1/k$ is the inverse of $k$ modulo $N$.  
Thus, one can see that 
\begin{align}\label{stackN}
    Z_{\widetilde{\cal PF}}[s; k] = 
    \omega^{k \text{Arf}_N(s)} Z_{\cal PF}[ ks' ; \frac1k ]\ . 
\end{align}
As expected, \eqref{stackN} reduces to \eqref{stack} for $N=2$ and $k=1$. 

To summarize, one has the following $\mathbb{Z}_N$ duality web \cite{Thorngren:2021yso} which generalizes the $\mathbb{Z}_2$ duality web:
\begin{center}\label{Tab:ZN}
        \begin{tikzpicture}[node distance = 6cm, thick]% 
        \node (B) {$\mathcal{B}$};
        \node (Bdual) [below= 3cm of B] {$\widetilde{\mathcal{B}} = \mathcal{B}/\mathbb{Z}_N$};
        \node (PF) [right=of B] {$\mathcal{PF}$};
        \node (PFdual) [below= 3cm of PF] {$\widetilde{\mathcal{PF}}$};
        \draw[<->] (B) -- node [midway,above] {Parafermionization} node [midway,below] {Bosonization} (PF);
        \draw[<->] (B) -- node [sloped,below]{Orbifolding} (Bdual);
        \draw[<->] (Bdual) -- node [midway,above] {Parafermionization} node [midway,below] {Bosonization} (PFdual);
        \draw[<->] (PF) -- (PFdual);
    \end{tikzpicture}
\end{center}

\subsection{Modular properties}

As mentioned earlier, it requires a careful analysis to understand 
how the torus partition function of a parafermionic theory transforms
under the modular group $SL(2,\mathbb{Z})$. Upon a modular transformation that 
maps a modulus $\tau$ to 
\begin{align}\label{SL(2,Z)}
    \tau' = \frac{a\tau + b}{ c\tau +d} \ , 
\end{align}
where $a,b,c$ and $d$ are integers and $ad-bc=1$, 
the basis one-cycles $\Gamma_I$ transform as 
\begin{align}\label{mapslgamma}
    \Gamma_1 & \mapsto  a \Gamma_1' - c \Gamma_2' \ , 
    \nonumber \\ 
    \Gamma_2 & \mapsto - b \Gamma_1' + d\Gamma_2' \ .
\end{align}
It implies that the modular transformation \eqref{SL(2,Z)} also acts on 
holonomies $(a_1,a_2)$ by 
\begin{align}
    a_1 & \mapsto  a a_1' - c a_2' \ , 
    \nonumber \\ 
    a_2 & \mapsto - b a_1' + d a_2' \ ,    
\end{align}
where $a_i'$ is a $\mathbb{Z}_N$ holonomy along $\Gamma_i'$. 

One can define the canonical topological boundary states with 
$\tau'$ as follows,
\begin{align}
    U[\Gamma_1'] | (a_1',a_2'):\tau'\rangle 
    & = \omega^{a_1'} | (a_1',a_2'):\tau'\rangle \ ,
    \nonumber\\ 
    U[\Gamma_2'] | (a_1',a_2'):\tau'\rangle 
    & = \omega^{a_2'} | (a_1',a_2'):\tau'\rangle \ .    
\end{align}
Here, we present the modulus dependence explicitly for clarity. 
We can then verify that both $|(a_1,a_2):\tau\rangle$ 
and $|(a_1',a_2'):\tau'\rangle$ with
\begin{align}\label{sl01}
    (a_1',a_2') = (d a_1 + ca_2, ba_1 + a a_2)
\end{align}
share the same eigenvalues of $U[\Gamma]$. To see this,    
we first decompose the cycle $\Gamma$ as
$\Gamma= m \Gamma_1 + n \Gamma_2$ for the modulus $\tau$
while as $\Gamma=m' \Gamma_1' + n'\Gamma_2'$ for $\tau'$ 
with
\begin{align}\label{sl02}
    m' = (ma - n b) \ , \quad 
    n' = (- mc + nd )\ .
\end{align}
The eigenvalues for $\tau'$ become 
\begin{align}
    U[\Gamma] | a' :\tau' \rangle  & = \omega^{ m' a_1' + n' a_2'} | a' :\tau' \rangle\ ,
    \nonumber \\ 
    & = \omega^{m a_1 + n a_2 } | a' :\tau' \rangle \ ,
\end{align}
where we used \eqref{sl01} and \eqref{sl02}. Namely, 
they are the same as those for $\tau$. One can thus conclude that
\begin{align}\label{sltrfora}
    \Big| (da_1+ca_2, ba_1+ aa_2) :\tau' = \frac{a\tau+b}{c\tau+d} \Big\rangle 
    = \Big| (a_1, a_2) : \tau\Big\rangle\ .     
\end{align}
In particular, under $T$ and $S$ transformations, 
\begin{align}\label{modularabasis}
    | (a_1,a_2):\tau \rangle & = | (a_1, a_1+a_2) : \tau+1 \rangle\ ,      
    \nonumber \\ 
    | (a_1, a_2) : \tau \rangle &= | (a_2, -a_1) : -1/\tau \rangle\ .  
\end{align}
Consequently, the BF model on the slab 
with a final state \eqref{modularabasis} shows that 
the bosonic partition functions transform covariantly under $SL(2,\mathbb{Z})$, 
\begin{align}
    Z_{\cal B}[a_1,a_2+a_1:\tau+1]  = Z_{\cal B}[a_1,a_2:\tau] = 
    Z_{\cal B}[a_2, -a_1: -1/\tau] \ .
\end{align}

Similarly, parafermionic 
topological boundary states with $\tau'$ \eqref{SL(2,Z)} 
can be defined as eigenstates of loop operators $U'_{Pf;k}[\Gamma]$,
\begin{align}
    U'_{Pf;k}[\Gamma_1'] | (s_1',s_2'):\tau' \rangle & = 
    \omega^{-k s_1'} | (s_1',s_2'):\tau' \rangle\ ,
    \nonumber \\ 
    U'_{Pf;k}[\Gamma_2'] | (s_1',s_2'):\tau' \rangle & = 
    \omega^{ k s_2'} | (s_1',s_2'):\tau' \rangle\ , 
\end{align}
where $U'_{Pf;k}[\Gamma]$ are loop operators obeying the 
constraints from \eqref{condition01} to \eqref{condition03}
with 
\begin{align}
    \epsilon(\gamma'_1, \gamma'_2) = \omega^{-k} \ .    
\end{align}
In other words, the symmetric pairing for $\tau'$ 
is chosen as $(\gamma_1',\gamma_2')'=-1$ and $(\gamma'_i,\gamma'_i)'=0$ for $i=1,2$. 
One can easily see that 
$\{ |(s_1,s_2):\tau \rangle \} $ and $\{| (s_1',s_2'):\tau' \rangle\}$
do not diagonalize the same operators. This is because 
$U_{Pf;k}[\Gamma]$ for the modulus $\tau$ and 
$U'_{Pf;k}[\Gamma]$ for the modulus $\tau'$ are simply not equivalent. 
As explained before, the above non-equivalence essentially boils down to the dependence 
of the symmetric pairing on a choice of basis for $N>2$.

The non-covariance of $U_{Pf;k}[\Gamma]$ under $SL(2,\mathbb{Z})$ for $N>2$ 
results in somewhat sophisticated modular properties of the 
parafermionic topological boundary states. To unravel them, it is convenient to  
begin by a representation of $| (s_1',s_2'):\tau' \rangle$ in the basis of $|a':\tau'\rangle $ for $\tau'$ \eqref{SL(2,Z)}, 
\begin{align}
    | s'; k :\tau' \rangle= \frac1N \sum_{a'} \omega^{-k \text{Arf}_N(s'+a')} 
    |a' :\tau' \rangle\ .
\end{align}
Upon rewriting it in terms of the topological boundary states for $\tau$ 
by means of \eqref{sltrfora} and the inverse of \eqref{eq:ZNs}, we can read off 
the modular transformation rules for parafermionic states 
\begin{align}
    | s' ; k :\tau' \rangle & = \frac1N \sum_{a} \omega^{-k \text{Arf}_N(s'+a')} |a:\tau \rangle\ ,
    \nonumber \\ & = 
    \frac{1}{N^2} \sum_{a,s}  \omega^{-k \text{Arf}_N(s'+a') + k \text{Arf}_N(s+a)} |s; k :\tau \rangle\ ,
\end{align}
where $(a_1',a_2') = ( da_1+ca_2, ba_1+ a a_2)$. Accordingly, we learn that 
\begin{align}
    Z_{\cal PF}\Big[s';k : \tau' = \frac{a\tau+b}{c\tau+d}\Big] & = \sum_{s} \rho(\Lambda)_{s_1',s_2'}^{s_1,s_2} ~ Z_{\cal PF}\Big[s;k:\tau\Big]\ , 
\end{align}
where the modular weight is zero and $\rho(\Lambda)$ is a representation of $\Lambda=\begin{pmatrix} a & b \\ c &d \end{pmatrix} \in SL(2,\mathbb{Z})$ in the space of parafermionic partition functions, 
\begin{align}\label{modularmatrices}
    \rho(\Lambda)_{s_1',s_2'}^{s_1,s_2} & = 
    \frac{1}{N^2} \sum_{a} (\omega^k)^{ \text{Arf}_N(s'+a') - \text{Arf}_N(s+a) }
\end{align}
with  $(a_1',a_2') = ( da_1+ca_2, ba_1+ a a_2)$. 

One may wonder if \eqref{modularmatrices} is a consistent representation of $SL(2,\mathbb{Z})$. 
To confirm it, we need to show that  \eqref{modularmatrices} satisfies two relations below 
\begin{align}
    \rho({\bf 1}) = {\bf 1}_{N\times N}\ , \quad 
    \rho(\Lambda_2) \rho(\Lambda_1) = \rho(\Lambda_2 \Lambda_1)\ .  
\end{align}
Since the former is evident, 
we only present elementary calculations to demonstrate
that \eqref{modularmatrices} obeys the latter. 
To this end, let us consider two elements, $\Lambda_1$ and 
$\Lambda_2$, of the modular group,
\begin{align}
    \Lambda_1 = 
    \begin{pmatrix} 
        a & b \\ c & d
    \end{pmatrix}\ , \quad 
    \Lambda_2 = 
    \begin{pmatrix} 
        e & f \\ g & h
    \end{pmatrix}\ . 
\end{align}
Then, one can easily manage to rewrite the product of $\rho(\Lambda_1)$ 
and $\rho(\Lambda_1)$ as 
\begin{align}\label{repcheck01}
    \rho(\Lambda_1)_{t_1,t_2}^{s_1',s_2'} \rho(\Lambda_2)_{s_1',s_2'}^{s_1,s_2} & = 
    \frac{1}{N^4} \sum_{a,b,s'} 
    (\omega^k)^{ \text{Arf}_N(t+b') + \text{Arf}_N(s'+a') - \text{Arf}_N(s'+b)- \text{Arf}_N(s+a) }
    \nonumber \\ & = 
    \frac{1}{N^2} \sum_{a} (\omega^k)^{\text{Arf}_N(t+ a'') - \text{Arf}_N(s+ a)}
\end{align}
where $(a_1', a_2')= (d a_1+ca_2, ba_1+a a_2)$, $(b_1',b_2')= (h b_1 + g b_2, f b_1+ e b_2)$, 
and 
\begin{align}
    \begin{pmatrix} a_1''\\ a_2''  \end{pmatrix} = 
    \begin{pmatrix} 
        h & g \\ f & e
    \end{pmatrix} 
    \begin{pmatrix} 
        d & c \\ b & a
    \end{pmatrix}    
    \begin{pmatrix} a_1\\ a_2  \end{pmatrix}\ .
\end{align}
Note that the sum over $s_1'$ and $s_2'$ in the first line of 
\eqref{repcheck01} identically vanishes unless 
$a_i' = b_i$ ($i=1,2$). The right hand side of \eqref{repcheck01} 
indeed agrees with $\rho(\Lambda_2 \Lambda_1)_{t_1,t_2}^{s_1,s_2}$, 
which ensures that \eqref{modularmatrices} is a representation of $SL(2,\mathbb{Z})$.
For instance, the above representation satisfies 
\begin{align}
    \big( \rho(S)^2 \big)_{t_1,t_2}^{s_1,s_2} = \big( (\rho(S) \rho(T))^3 \big)_{t_1,t_2}^{s_1,s_2} = \delta^{s_1}_{-t_1} \delta^{s_2}_{-t_2} \ . 
\end{align}
We remark that \eqref{modularmatrices} for $k=1$ matches with the modular matrix studied in \cite{Yao:2020dqx}, and reduces to that of fermionic partition functions
when $N=2$. As a final comment, for fermionic theories each sector is (at least) preserved by a certain index-2 subgroup of $SL(2,\mathbb{Z})$, which imposes strong constraints on the torus partition function \cite{Bae:2020xzl,Bae:2021mej,Duan:2022kxr}. For general parafermionic theories the modular transformation is much more complicated, and it would be interesting to study its constraint on the spectrum.

%%%%%%%%%%%%%%%%%%%%%%%%%%%%%%%%%%%%%%%%%%%%%%%%%%%%%%%%%%%%%%%%%%%%%%%%%%%%%%%%%%%%%%%%%%%%%%%%%%%%%%%%%%%%%%%%%%%%%%%%%
%%%%%%%%%%%%%%%%%%%%%%%%%%%%%%%%%%%%%%%%%%%%%%%%%%%%%%%%%%%%%%%%%%%%%%%%%%%%%%%%%%%%%%%%%%%%%%%%%%%%%%%%%%%%%%%%%%%%%%%%%
%%%%%%%%%%%%%%%%%%%%%%%%%%%%%%%%%%%%%%%%%%%%%%%%%%%%%%%%%%%%%%%%%%%%%%%%%%%%%%%%%%%%%%%%%%%%%%%%%%%%%%%%%%%%%%%%%%%%%%%%%

\subsection{Partial gauging}

When $N$ is not a prime number, one can partially gauge 
a subgroup of $\mathbb{Z}_N$. Without loss of generality, let $N=P Q$ where $P$ and $Q$ 
need not be relatively prime in general. Based on 
the idea of the Symmetry TFT, we discuss in this section 
the partial orbifold ${\cal B}/\mathbb{Z}_P$ 
and $\mathbb{Z}_P$ parafermionization. We also 
argue that they have mixed 't Hooft anomaly unless $P$ and $Q$ 
are co-prime. 

For the partial orbifold, we need to construct a basis of the Hilbert space of the BF model on $\Sigma_g$
in which $\{ U^P[\Gamma] \} $ and $\{ \widetilde{U}^Q[\Gamma] \} $ are diagonalized. 
Note that the operator $U^P[\Gamma]$ generates $\mathbb{Z}_Q$ subgroup of $\mathbb{Z}_N$
while $\widetilde{U}^Q[\Gamma]$ generates $\mathbb{Z}_P$. 
The basis of our interest can thus be described by $\mathbb{Z}_Q$-, and 
$\mathbb{Z}_P$-valued holonomies. 

To construct such eigenstates, 
we first decompose the $\mathbb{Z}_N$-valued 
holonomy $a$ as
\begin{align}
    | b,c \rangle \equiv |a=b + Q c \rangle \ ,     
\end{align}
where the holonomy $c$ is guaranteed to be $\mathbb{Z}_P$-valued
\begin{align}
    | b , c+ P \gamma_i\rangle = | b , c\rangle \ .
\end{align}
Here $\gamma_i$ are again unit vectors in the $N^{2g}$-dimensional space of 
holonomy $a$, dual to basis one-cycles $\Gamma_i$ of the given Riemann surface. 
On the other hand, one can see 
\begin{align}
    |b + Q \gamma_i, c \rangle = | b , c + \gamma_i \rangle \ .
\end{align}

We propose that 
the partial $\mathbb{Z}_P$ orbifold can be described by topological boundary states at $t=1$ defined as 
\begin{align}\label{Zporbifold}
    | b , \tilde{c}  \rangle = \frac{1}{P^g} \sum_{c} (\omega^Q)^{ \langle c, \tilde{c} \rangle }  | b , c\rangle.  
\end{align}
By definition, the state \eqref{Zporbifold} remains unchanged under the shift
$\tilde c \to \tilde c + P \gamma_i$. Moreover, the holonomy variable $b$ now becomes $\mathbb{Z}_Q$-valued effectively.
This is because two states whose difference is the shift of $b$ by $Q\gamma_i$
essentially describe the same quantum state,  
\begin{align}\label{anomaly01}
| b + Q \gamma_i, \tilde{c} \rangle = (\omega^Q)^{-\langle \gamma_i,\tilde c \rangle } | b , \tilde{c} \rangle\ .
\end{align}
It implies that the orbifold ${\cal B}/\mathbb{Z}_P$ has $\mathbb{Z}_Q \times \widetilde{\mathbb{Z}}_P$ symmetry
whose holonomies can be identified as $b$ and $\tilde c$. 
They indeed diagonalize the very operators with eigenvalues as follows,  
\begin{align}
    U^P[\Gamma] | b , \tilde{c}\rangle & = \frac{1}{P^g} \sum_c \omega^{Q\langle \tilde c,c\rangle} (\omega^P)^{\langle \gamma, b + Q c \rangle } | b,c\rangle   \ ,
    \nonumber \\ & = 
    (\omega^P)^{\langle \gamma, b  \rangle } | b, \tilde{c}\rangle\ , 
\end{align}
and
\begin{align}
    \widetilde{U}^Q[\Gamma] | b , \tilde{c}\rangle & =  
%    \frac{1}{P^g} \sum_c \omega^{Q\langle \tilde c,c\rangle}  \widetilde{U}^Q[\Gamma] | b,c \rangle  
%    \nonumber \\ & = 
    \frac{1}{P^g} \sum_c \omega^{Q\langle \tilde c,c\rangle} | b , c- \gamma \rangle 
    \nonumber \\ & = 
    (\omega^Q)^{\langle \gamma, \tilde c \rangle} | b, \tilde c \rangle \ .
\end{align}
Note that $\omega^P$ and $\omega^Q$ are $Q$-th and $P$-th roots of unity. 

When a final state is chosen by \eqref{Zporbifold}, the BF model 
on the slab gives the partition function of the orbifold ${\cal B}/\mathbb{Z}_P$
with both $\mathbb{Z}_Q$ and $\widetilde{\mathbb{Z}}_P$ weakly gauged, 
\begin{align}
    Z_{{\cal B}/\mathbb{Z}_P}[b,\tilde c] = \langle b, \tilde c |\chi \rangle 
    = \frac{1}{P^g} \sum_{c} \omega^{Q\langle \tilde c , c\rangle } Z_{\cal B}[ b + Q c].    
\end{align}
The non-trivial phase of \eqref{anomaly01} suggests that 
the orbifold ${\cal B}/\mathbb{Z}_P$  has an anomalous global symmetry.
Specifically, under a large gauge transformation $b \to b + Q\gamma_i$, 
the partition function $Z_{{\cal B}/\mathbb{Z}_P}$ fails to be invariant but  
\begin{align}\label{Partial-gauging-anomalous-phase}
     Z_{{\cal B}/\mathbb{Z}_P}[b + Q\gamma_i,\tilde c] = \omega^{Q \langle \gamma_i, \tilde c \rangle }
    Z_{{\cal B}/\mathbb{Z}_P}[b ,\tilde c]
\end{align}
Since there exists a nontrivial phase only after we gauge 
the group $\widetilde{\mathbb{Z}}_P$, the orbifold ${\cal B}/\mathbb{Z}_P$ has a mixed 't Hooft anomaly. 
We shall discuss in detail that the above anomalous phase agrees with 
the results in \cite{Bhardwaj:2017xup} shortly. 

When $P$ and $Q$ are relatively prime, the anomalous phase can be removed 
by alternative decomposition of $\mathbb{Z}_N$ holonomy $a$, 
\begin{align}
    |b' , c\rangle \equiv | a = P b' + Q c \rangle\ .
\end{align}
Since $\mathbb{Z}_N$ is equivalent to $\mathbb{Z}_P \times \mathbb{Z}_Q$ due to the Chinese remainder theorem, 
one can unambiguously regard $b'$ and $c$ as $\mathbb{Z}_Q$- and $\mathbb{Z}_P$-holonomies, 
\begin{align}
    | b' + Q \gamma_i, c + P \gamma_j \rangle = |b',c \rangle,     
\end{align}
for any $\gamma_i$ and $\gamma_j$. As a consequence, the partition function 
of the orbifold ${\cal B}/\mathbb{Z}_P$ becomes invariant under the 
large gauge transformation, 
\begin{align}
    Z_{{\cal B}/\mathbb{Z}_P}[b'+ Q\gamma_i,\tilde c+ P \gamma_j] 
    = Z_{{\cal B}/\mathbb{Z}_P}[b',\tilde c]\ .
\end{align}

Parallel to \eqref{Zporbifold}, we also propose that topological boundary states 
for the partial $\mathbb{Z}_P$ parafermionization are given by
\begin{align}\label{Zppara}
    |b, s ; l \rangle \equiv   \frac1P  \sum_{c} (\omega^Q)^{- l \text{Arf}_P (s + c)  } | b,c  \rangle\ ,
\end{align}
where $l$ is co-prime with $P$. Since 
\eqref{Zppara} is invariant under the shift $s \to s+ P \gamma_i$, 
we can regard $s$ as the $\mathbb{Z}_P$ paraspin structures. 
However, the holonomy $b$ is not $\mathbb{Z}_Q$-valued 
\begin{align}\label{Zpparamap}
    |b + Q\gamma_i, s ; l\rangle & = \sum_{c} (\omega^Q)^{-l \text{Arf}_P (s + c-\gamma_i)} | b, c\rangle 
    \nonumber \\ & = | b, s-\gamma_i; l \rangle\ . 
\end{align}
The state defined by \eqref{Zppara} is an eigenstate of a loop operator 
\begin{align}\label{eq:PFoperator_partial}
    U_{Pf;l}^{(P)}[\Gamma]  = U[m\Gamma_1]^{-l} U[n\Gamma_2]^{l} \widetilde{U}[m\Gamma_1]^Q 
    \widetilde{U}[n\Gamma_2]^Q 
\end{align}
with the eigenvalue  
\begin{align}
    U_{Pf;l}^{(P)}[\Gamma] | b ,s ; l \rangle & = \frac1P \sum_c \omega^{-l \text{Arf}_P(s+c+\gamma)} 
    \omega^{l (\gamma, b+ Q c )} |b, c \rangle\ , 
    \nonumber \\ & = \omega^{l(\gamma,b)} \cdot 
    (\omega^{Q})^{- l (\text{Arf}_P (s+\gamma) - \text{Arf}_P(s) )} | b, s; l\rangle\ .  
\end{align}
Based on \eqref{quadrefinement03}, the exponent 
$Q^{(P)}_s(\gamma) \equiv \big( \text{Arf}_P(s+\gamma) - \text{Arf}_P(s)\big)$ 
can be identified as a quadratic refinement of the $\mathbb{Z}_P$ symmetric pairing. 
Since the loop operators $U_{Pf;l}^{(P)}(\Gamma)$ meet the conditions from 
\eqref{condition01} to \eqref{condition03} with 
\begin{align}
    \epsilon(\gamma,\gamma') = (\omega^Q)^{l (\gamma, \gamma')}\ , 
\end{align}
\eqref{Zppara} can be understood 
as the topological boundary states for the partial parafermionization. 
Accordingly, the torus partition function of the $\mathbb{Z}_P$ 
parafermionic theory becomes
\begin{align}
    Z_{{\cal PF}}^{(P)}[b,s;l] 
    = \langle b,s ; l | \chi \rangle = 
    \frac1P \sum_c (\omega^Q)^{l \text{Arf}_{P}(s+c) } Z_{\cal B}[b+Qc]\ .  
\end{align}
%

%%%%%%%%%%%%%%%%%%%%%%%%%%%%%%%%%%%%%%%%%%%%%%%%%%%%%%%%%%%%%%%%%%%%%%%%%%%%%%%%%%%%%%%%
%%%%%%%%%%%%%%%%%%%%%%%%%%%%%%%%%%%%%%%%%%%%%%%%%%%%%%%%%%%%%%%%%%%%%%%%%%%%%%%%%%%%%%%%
%%%%%%%%%%%%%%%%%%%%%%%%%%%%%%%%%%%%%%%%%%%%%%%%%%%%%%%%%%%%%%%%%%%%%%%%%%%%%%%%%%%%%%%%

\subsection{'t Hooft anomaly revisited}

As studied in \cite{Bhardwaj:2017xup}, the 't Hooft anomaly is encoded 
in the so-called `associator' of a network of symmetry defects. 
We argue in this section that the anomalous phase observed in  
\eqref{Partial-gauging-anomalous-phase} is consistent with 
the conventional anomalous phase in the defect network. 

\begin{figure}[t!]
        \centering
        \includegraphics[scale=0.55]{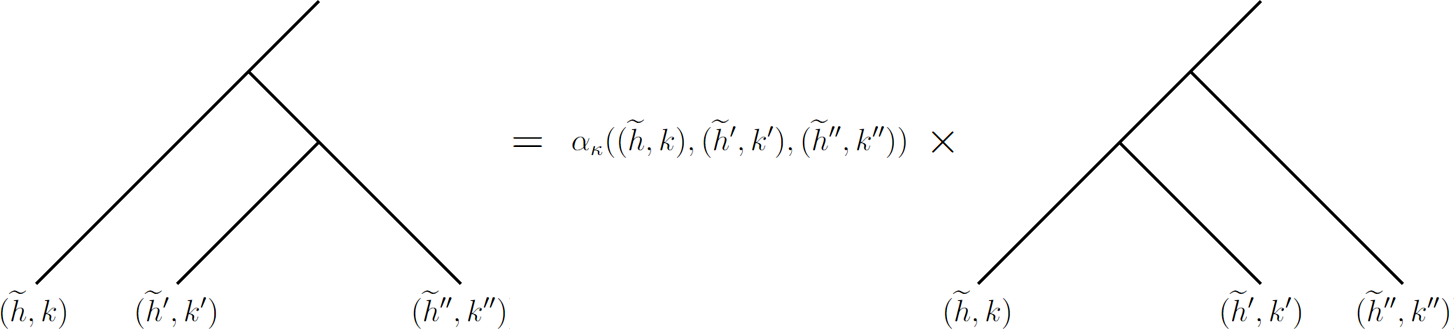}
        \caption{F-move in the dual theory $\widetilde{G}$.}
        \label{fig:Partial-gauging-F-move}
\end{figure}
We begin by a statement in \cite{Bhardwaj:2017xup} relevant to our discussion,
\begin{statement}
    Consider $H$ as a normal Abelian subgroup of $G$ and denote $K = G/H$. A group element $g \in G$ can be decomposed as a pair $(h,k) \in H \times K$ such that the group structure is given by,
        \begin{equation}
            (h,k) (h',k') = (h k h' k^{-1} \kappa(k,k') , k k'),
      \end{equation}
    where $\kappa(k,k')$ is an $H$-valued 2-cocycle of $K$. We will write $G = H \rtimes_{\kappa} K$.

    Then we gauge $H$ and denote by $\widetilde{H}$ the dual group of $H$. $K$ remains unchanged and the whole symmetry group is $\widetilde{G} = \widetilde{H} \rtimes K$ where $\widetilde{\kappa}$ is trivial. $\widetilde{G}$ has a 't Hooft anomaly and the associator $\alpha_{\kappa}$ (see figure \ref{fig:Partial-gauging-F-move}) is induced by $\kappa$ as,
        \begin{equation}\label{Partial-gauging-associator}
            \alpha_{\kappa}((\widetilde{h},k),(\widetilde{h}',k'),(\widetilde{h}'',k'')) = \widetilde{h}'' ( (k k') \kappa(k,k') (k k')^{-1}),
        \end{equation}
    where $\widetilde{h},\widetilde{h}',\widetilde{h}''\in \widetilde{H}$ and they have a natural action on $H$. 
\end{statement}

\begin{figure}[t!]
    \centering
    \includegraphics[scale=0.6]{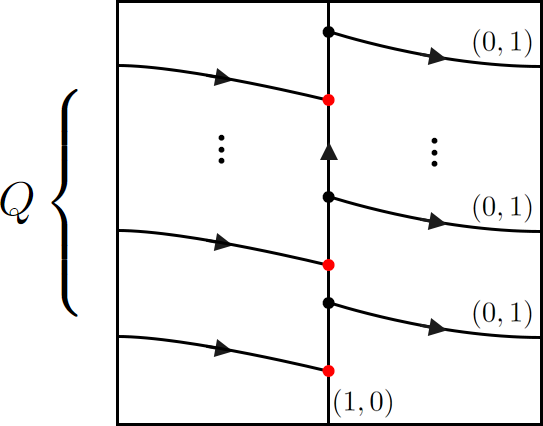}
    \caption{The network corresponding to $Z_{\mathcal{B}/\mathbb{Z}_P}(b = -Q \gamma_1 , \widetilde{c} = -\gamma_2)$. }
    \label{fig:Partial-gauging-defect-network}
\end{figure}

In the present case, $G= \mathbb{Z}_{PQ}$, $H = \mathbb{Z}_P$ and $K=\mathbb{Z}_Q$. Let us assume $P,Q$ are not coprime. For any $g$ in $\mathbb{Z}_{PQ}$, one can decompose it as $g = k + Q h$ with $h \in \mathbb{Z}_{P}$ and $k \in \mathbb{Z}_Q$. The 2-cocycle $\kappa$ is determined by the group law of $G$,
\begin{equation}
    \kappa(k,k') = \left\{ \begin{array}{c}
         0\quad (k+k'<Q)  \\
         1\quad (k+k' \geq Q) 
    \end{array}\right.
\end{equation}
After gauging $\mathbb{Z}_P$, the symmetry group is $\widetilde{G} = \widetilde{\mathbb{Z}}_P \rtimes \mathbb{Z}_Q$ with a 't Hooft anomaly given by \eqref{Partial-gauging-associator} and we will denote the group element as a pair $(\widetilde{h},k)$ with $\widetilde{h}\in \widetilde{Z}_P$ and $k \in \mathbb{Z}_Q$. To illustrate the anomalous phase \eqref{Partial-gauging-anomalous-phase} of the torus partition function $Z_{\mathcal{B}/\mathbb{Z}_P}$, we turn on a $\widetilde{\mathbb{Z}}_P$ background by inserting a $(1,0)$ defect along the time direction. Consider the configuration shown in figure \ref{fig:Partial-gauging-defect-network} where we insert $Q$ $(0,1)$-operators stretching along the spatial direction. The partition function represented by this network is $Z_{\mathcal{B}/\mathbb{Z}_P}(b = -Q \gamma_1 , \widetilde{c} = -\gamma_2)$.

\begin{figure}[t!]
    \centering
    \includegraphics[scale=0.6]{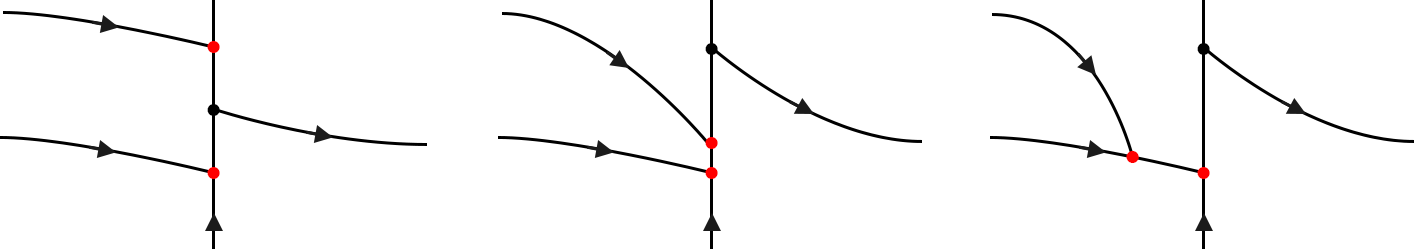}
    \caption{Resolve the network by moving red nodes.}
    \label{fig:Partial-gauging-node-corssing}
\end{figure}

%To resolve this network, let us first move all red nodes to the bottom along the vertical defect.
Let us apply various F moves to resolve this network. 
When a node crosses another node as shown in \ref{fig:Partial-gauging-node-corssing}, the partition function will develop a phase according to \eqref{Partial-gauging-associator}.
First, when the red node crosses the black node, the phase is trivial because the $\widetilde{h}''$ in \eqref{Partial-gauging-associator} is zero and we can move all red nodes across the black nodes. Then we can move the top $Q-1$ red nodes across the bottom red node one by one and attach them to the bottom horizontal operators as shown in the last diagram in figure \ref{fig:Partial-gauging-node-corssing}. The anomalous phase shows up when we move the last red node such that the triplet $((\widetilde{h},k),(\widetilde{h}',k'),(\widetilde{h}'',k''))$ is $((0,1),(0,Q-1),(1,0))$ and the associator $\alpha_{\kappa}$ is,
\begin{equation}
    \alpha_{\kappa}((0,1),(0,Q-1),(1,0)) = \omega^Q.
\end{equation}

\begin{figure}[t!]
    \centering
    \includegraphics[scale=0.6]{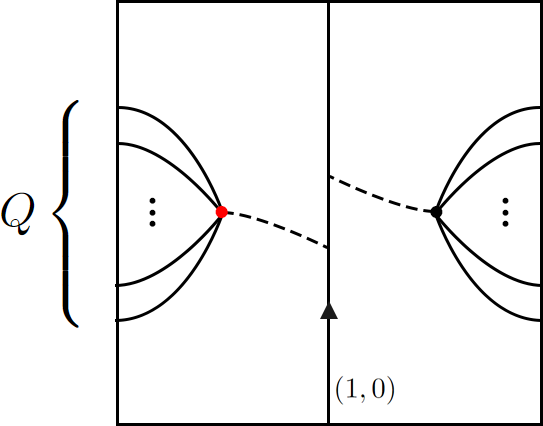}
    \caption{The network corresponding to $Z_{\mathcal{B}/\mathbb{Z}_P}(b = 0 , \widetilde{c} = -\gamma_2)$. }
    \label{fig:Partial-gauging-defect-network-final}
\end{figure}

We then apply the same procedure to the remaining black nodes and there are no additional phases because the $\widetilde{h}''$ is always zero during all the F-moves. Eventually, the network looks like figure \ref{fig:Partial-gauging-defect-network-final} where the dash lines are identity $(0,0)$. Since all defects are invertible, the bubble can shrink to nothing and the only defect remaining is the vertical $(1,0)$-defect. The partition function represented by this is simply $Z_{\mathcal{B}/\mathbb{Z}_P}(b = 0 , \widetilde{c} = -\gamma_1)$.

Therefore we have shown that,
\begin{equation}
    Z_{\mathcal{B}/\mathbb{Z}_P}(b = -Q \gamma_1 , \widetilde{c} = -\gamma_2) = \omega^Q Z_{\mathcal{B}/\mathbb{Z}_P}(b = 0 , \widetilde{c} = -\gamma_2),
\end{equation}
which is consistent with \eqref{Partial-gauging-anomalous-phase} using $\langle \gamma_1 , \gamma_2 \rangle = 1$.

%%%%%%%%%%%%%%%%%%%%%%%%%%%%%%%%%%%%%%%%%%%%%%%%%%%%%%%%%%%%%%%%%%%%%%%%%%%%%%%%%%%%%%%%%%%%%%%%%%%%%%%%%%%%%%%%%%%%%%
%%%%%%%%%%%%%%%%%%%%%%%%%%%%%%%%%%%%%%%%%%%%%%%%%%%%%%%%%%%%%%%%%%%%%%%%%%%%%%%%%%%%%%%%%%%%%%%%%%%%%%%%%%%%%%%%%%%%%%
%%%%%%%%%%%%%%%%%%%%%%%%%%%%%%%%%%%%%%%%%%%%%%%%%%%%%%%%%%%%%%%%%%%%%%%%%%%%%%%%%%%%%%%%%%%%%%%%%%%%%%%%%%%%%%%%%%%%%%

\section{Example: $SU(2)_N/U(1)_{2N}$ coset CFTs}\label{sec:4}

We demonstrate the $\mathbb{Z}_N$ orbifold and parafermionization
by examining a well-studied class of coset models with $\mathbb{Z}_N$ symmetry,
\begin{align}\label{eq:fields}
    {\cal B} = \frac{su(2)_N}{u(1)_{2N}}\ .  
\end{align}
The coset model has the central charge $c=2(N-1)/(N+2)$. Recently, it was shown in 
\cite{Lin:2021udi} that the $\mathbb{Z}_N$ symmetry of the model is non-anomalous. 
For $N = 2$, \eqref{eq:fields} can be identified as the Ising model, while for $N = 3$ it coincides with the three-state Potts model.

The model \eqref{eq:fields} has the Hilbert space that decomposes into
finitely many modules of its chiral algebra. For a given $N$, the chiral 
algebra allows chiral primaries of weight 
\begin{align}\label{conformalweight}
    h_{l,m} = \frac{l(l+2)}{4(N+2)} - \frac{m^2}{4N^2}\ . 
\end{align}
where $0\leq l \leq N$ and $|m|\leq l$. 
The torus partition function of the coset model can be described as 
\begin{align}\label{parafermionpartitionsum}
    Z_{\cal B} = \sum_{l,m}\Big| \chi_{l,m}(\tau) \Big|^2\ , 
\end{align}
where $\chi_{l,m}(\tau)$ refers to the conformal character for the chiral primary
of weight $h_{l,m}$ \eqref{conformalweight}. Since the chiral 
primaries are subject to various identification 
\begin{gather}
    \chi_{l,m}(\tau)= \chi_{l,-m}(\tau) = \chi_{l,m+2N}(\tau) = \chi_{N-l,N+m}(\tau)\ ,     
    \nonumber \\ 
    \chi_{l,m}(\tau) = 0 \quad  \text{ if } \quad m- l \in 2 \mathbb{Z}\ ,  
\end{gather}
the summation in \eqref{parafermionpartitionsum} runs over the set
\begin{align}
    \{ (l, m) | 0 \leq l \leq N, -l+2 \leq m \leq l, l-m  \in 2 \mathbb{Z} \}\ . 
\end{align}
In other words, the coset model ${\cal B}$ has $N(N+1)/2$ independent primary operators.

Turning on the background $\mathbb{Z}_N$ gauge field $a=(a_1,a_2)$, the partition function $Z_{\cal B}$
becomes 
\begin{align}\label{sdf}
    Z_{\mathcal{B}}[a_1,a_2] = \frac{1}{2} \sum_{l=0}^N \sum_{m=-N+1}^N  \omega^{(m+a_1)a_2} 
    \chi_{l,m+2a_1}(\tau) \chi^*_{l,m}(\bar \tau)\ .
\end{align}
Here, we used the $\mathbb{Z}_N$ action on each character 
\begin{align}
    \chi_{l,m}(\tau) \to e^{\frac{2\pi i m}{N}}  \chi_{l,m}(\tau)\ , 
\end{align}
and the explicit modular $S$-matrix 
\begin{align}
    S_{l,m ; k,n} = \frac{2}{\sqrt{N(N+2)}} e^{2\pi i \frac{m n}{2N}}
    \sin{\Big( \pi \frac{(l+1)(k+1)}{N+2} \Big)}\ , 
\end{align}
to express the twisted partition function $Z_{\cal B}[a_1,a_2]$ 
into the above simple form \cite{}. We can also see that 
$T$ and $S$ transformations rotate the twisted partition function
in a covariant manner. More precisely, under $T$-transformation, 
one can show that
\begin{align}
    Z_{\cal B}[a_1,a_2+a_1:\tau+1] & = 
    \frac{1}{2} \sum_{l=0}^N \sum_{m=-N+1}^N  \omega^{(m+a_1)(a_2+a_1)} \omega^{-a_1(m+a_1)} \chi_{l,m+2a_1}(\tau) \chi^*_{l,m}(\bar \tau)\ ,
    \nonumber \\ & = 
    Z_{\cal B}[a_1,a_2:\tau]\ . 
\end{align}
On the other hand, the $S$-transformation of 
the twisted partition function is 
\begin{align}
    Z_{\cal B}[a_2,-a_1: -1/\tau] = \frac12
    \sum_{l',l''=0}^N \sum_{m',m''=-N+1}^N
    c_{l',m';l'',m''}
    \chi_{l',m'}(\tau) \chi^*_{l'';m''}(\bar \tau)  
\end{align}
with
\begin{align}
    c_{l',m';l'',m''} & = \sum_{l=0}^N \sum_{m=-N+1}^N 
    \frac{\omega^{-(m+a_2)a_1 + \frac{(m+2a_2)m'-m m''}{2}}}{N(N+2)}  \sin \frac{\pi(l+1)(l'+1)}{N+2}  \sin \frac{\pi(l+1)(l''+1)}{N+2}
    \nonumber\ .
\end{align}
Since summing over $l$ forces $l'=l''$ while the sum over $m$ 
vanishes unless $m'=m''+2a_1$, one obtains
\begin{align}
    c_{l',m';l'',m''} = \omega^{-a_1a_2 + a_2 (m''+2a1) } \delta_{l',l''}
    \delta_{m',m''+2a_1}\ ,
\end{align}
and thus 
\begin{align}
    Z_{\cal B}[a_2,-a_1: -1/\tau] = Z_{\cal B}[a_1,a_2 :\tau]\ . 
\end{align}

It is well-known that the Ising model is
self-dual under the Krammers-Wannier duality. 
Similarly, the $\mathbb{Z}_N$ gauging maps 
the coset model \eqref{eq:fields} to itself 
\begin{align}
    \widetilde{\cal B} \equiv {\cal B}/\mathbb{Z}_N \simeq {\cal B}\ .     
\end{align}
To see this, let us recall that the partition 
function of the orbifold $\widetilde{\cal B}$ coupled to the background field 
is 
\begin{align}
    Z_{{\cal B}/\mathbb{Z}_N} [\tilde a_1, \tilde a_2] & = 
    \frac1N \sum_{a_1,a_2} \omega^{\tilde a_1 a_2 -\tilde a_2 a_1 } Z_{\cal B}[a_1,a_2]
    \\ & = 
    \frac{1}{2N} \sum_{l=0}^N \sum_{m=-N+1}^N  \sum_{a_1,a_2} \omega^{(m+a_1)a_2 - (a_1 \tilde{a}_2 - a_2 \tilde{a}_1)} \chi_{l,m+2a_1}(\tau) \chi^*_{l,m}(\bar \tau)\ ,
    \nonumber  
\end{align}
where we used \eqref{sdf} for the second equality.
Since the sum over the $\mathbb{Z}_N$ holonomy $a_2$ is identically zero 
unless $a_1= - (\tilde a_1 + m)$, the above expression can be simplified as 
follows 
\begin{align}
    Z_{{\cal B}/\mathbb{Z}_N} [\tilde a_1, \tilde a_2]  & = 
    \frac12 \sum_{l=0}^N \sum_{m=-N+1}^N  \omega^{(\tilde a_1 + m) \tilde a_2 } 
    \chi_{l,- m- 2 \tilde{a}_1}(\tau) \chi^*_{l,m}(\bar \tau)\ . 
\end{align}
It perfectly matches with $Z_{\cal B}[a]$ with $ (a_1,a_2) = (\tilde a_1,\tilde a_2)$
due to the charge conjugation symmetry $\chi_{l,m}(\tau)=\chi_{l,-m}(\tau)$.  

Let us finally move onto the parafermionic theories dual to the coset models. 
Based on the parafermionization  \eqref{ZN-parafermionization}, 
one can obtain the torus partition function of our interest
\begin{align}
    Z_{\cal PF}[s_1,s_2; k] & = 
    \frac1N \sum_{a_1,a_2} \omega^{k \text{Arf}_N (s+a)} Z_{\cal B}[a_1,a_2] \ , 
    \\ & = 
    \frac{1}{2N} \sum_{l=0}^N \sum_{m=-N+1}^N  \sum_{a_1,a_2} \omega^{k(s_1+a_1)(s_2+a_2)+(m+a_1)a_2} \chi_{l,m+2a_1}(\tau) \chi^*_{l,m}(\bar \tau)\ , \nonumber
\end{align}
where $Z_{\cal B}[a_1,a_2]$ is given by \eqref{sdf}. It can be further 
massaged into a simpler expression 
\begin{align}\label{ghj}
    Z_{\cal PF}[s_1,s_2; k] = \frac12 \sum_{l=0}^N \sum_{a_1} 
    \omega^{k (s_1+a_1)s_2 }   \Big[ &
    \chi_{l,ks_1 + (k-1) a_1}(\tau) \chi^*_{l, k s_1+ (k+1)a_1}(\bar \tau)
    \nonumber \\ +  &
    \chi_{l,ks_1 + (k-1) a_1+N}(\tau) \chi^*_{l, k s_1+ (k+1)a_1+N}(\bar \tau)
    \Big]\ .
\end{align}
Note that, for $k=1$, \eqref{ghj} reduces to the torus partition 
function proposed in  \cite{Yao:2020dqx}. In this case the $\mathbb{Z}_N$ duality web in Table \ref{Tab:ZN} follows from the identity
\begin{align}
Z_{\cal PF}[-k s_1, ks_2; 1/k] = \omega^{-k s_1 s_2} Z_{\cal PF}[s_1, s_2; k]\ .
\end{align}

\subsection*{Acknowledgements}
We would like to thank Chi-Ming Chang, Jin Chen and Kimyeong Lee for helpful discussions. We also thank Chi-Ming Chang for comments on the draft. ZD thanks Fudan University, Shanghai for hospitality where part of this project was carried out. The main result of this work was presented by QJ at the 4th National Workshop on Fields and Strings held in Nanjing, and QJ
is grateful to the audience for feedback. ZD, QJ, and SL are supported by KIAS
Grant PG076902, PG080802 and PG056502.

\newpage

\appendix

\section{Operators defining parafermionic states}\label{App:operators}

In this appendix, we will give a systematic discussion about operators $U_{Pf;k}[\Gamma]$ defined in \eqref{ZN-exotic-operators-UPFp}. We begin with the general sets of operators labeled by a pair $(k,l)$,
\begin{equation}
    U_{Pf;k,l}[\Gamma] = U[m\Gamma_1]^{-k} U[n\Gamma_2]^k \widetilde{U}[m\Gamma_1]^l \widetilde{U}[n\Gamma_2]^l,
\end{equation}
where the operators in \eqref{ZN-exotic-operators-UPFp} corresponds to $l=1$. They satisfy,
\begin{align}
    &U_{Pf;k,l}[\Gamma] U_{Pf;k,l}[\Gamma'] \nonumber\\
    =& U[m\Gamma_1]^{-k} U[n\Gamma_2]^k \widetilde{U}[m\Gamma_1]^l \widetilde{U}[n\Gamma_2]^l U[m'\Gamma_1]^{-k} U[n'\Gamma_2]^k \widetilde{U}[m'\Gamma_1]^l \widetilde{U}[n'\Gamma_2]\nonumber\\
    =&U[(m+m')\Gamma_1]^{-k} U[(n+n')\Gamma_2]^k \widetilde{U}[(m+m')\Gamma_1]^l \widetilde{U}[(n+n')\Gamma_2]^l  \omega^{kl\langle m\gamma_1+n\gamma_2, m'\gamma_1-n'\gamma_2\rangle}\nonumber\\
    =&U_{Pf;k,l}[\Gamma+\Gamma'] \omega^{kl (\gamma,\gamma')},
\end{align}
where $(*,*)$ is the symmetric pairing defined as $(a,b) = a_1 b_2 + a_2 b_1$. The operators labeled by the same pair $(k,l)$ commute with each other. For operators with different $(k,l)$-pair, one has the commutation relation,
\begin{align}\label{eq:commutation}
    &U_{Pf;l,k}[\Gamma] U_{Pf;l',k'}[\Gamma'] \nonumber\\
    =& U[m\Gamma_1]^{-k} U[n\Gamma_2]^k \widetilde{U}[m\Gamma_1]^l \widetilde{U}[n\Gamma_2]^l U[m'\Gamma_1]^{-k'} U[n'\Gamma_2]^{k'} \widetilde{U}[m'\Gamma_1]^{l'}\widetilde{U}[n'\Gamma_2]^{l'}\nonumber\\
    =& U[m'\Gamma_1]^{-k'} U[n'\Gamma_2]^{k'} \left(U[m\Gamma_1]^{-k} U[n\Gamma_2]^k \widetilde{U}[m\Gamma_1]^l \widetilde{U}[n\Gamma_2]^l\right)\widetilde{U}[m'\Gamma_1]^{l'}\widetilde{U}[n'\Gamma_2]^{l'} \omega^{lk'(\gamma,\gamma')}\nonumber\\
    =&U[m'\Gamma_1]^{-k'} U[n'\Gamma_2]^{k'} \widetilde{U}[m'\Gamma_1]^{l'}\widetilde{U}[n'\Gamma_2]^{l'} U[m\Gamma_1]^{-k} U[n\Gamma_2]^k \widetilde{U}[m\Gamma_1]^l \widetilde{U}[n\Gamma_2]^l \omega^{(lk'-l'k)(\gamma,\gamma')} \nonumber\\
    =&U_{Pf;l',k'}[\Gamma'] U_{Pf;l,k}[\Gamma]\omega^{(lk'-l'k)(\gamma,\gamma')}.
\end{align}
They are commuting only if $lk'-l'k=0$, which is equivalent to $k'=\lambda k, l'=\lambda l$ for some non-zero $\lambda$. If $\lambda$ is coprime with $N$, the operators $U_{Pf;k',l'}$ is equivalent to $U_{Pf;k,l}$ up to rescaling of $\Gamma$, otherwise it is a subalgebra of $U_{Pf;k,l}$.

There are three cases depending on whether $k$ and $l$ are coprime with $N$:
\begin{itemize}
    \item If $l$ is coprime with $N$, then one can choose a  representative $l=1$ up to rescaling of $\Gamma$,
\begin{equation}
    U_{Pf;k,1}[\Gamma] = U[m\Gamma_1]^{-k} U[n\Gamma_2]^k \widetilde{U}[m\Gamma_1] \widetilde{U}[n\Gamma_2],
\end{equation}
which is the operators defined in \eqref{ZN-exotic-operators-UPFp}.
    \item If $k$ is coprime with $N$, then we can set $k=1$ and get,
\begin{equation}
     U_{Pf;1,l}[\Gamma]  = U[m\Gamma_1]^{-1} U[n\Gamma_2] \widetilde{U}[m\Gamma_1]^l 
    \widetilde{U}[n\Gamma_2]^l. 
\end{equation}
This kind of operator is related to the parafermionic operator \eqref{eq:PFoperator_partial} when we discuss partial gauging.
    \item Both $k$ and $l$ are not coprime with $N$. We did not consider this case in the paper so we omit the discussion here.
\end{itemize}

%\begin{thebibliography}{99}

%\end{thebibliography}
\bibliographystyle{JHEP}
\bibliography{main}

\end{document}